\documentclass[aps,prl,twocolumn,floatfix]{revtex4}
\pdfoutput=1
\usepackage[caption=false]{subfig}
\usepackage{graphicx}

\begin{document}

\title{Landau-Zener-St\"uckelberg-Majorana interferometry
of a single hole}

\author{Alex Bogan} 
\affiliation{Emerging Technology Division,  
National Research Council, Ottawa, Canada, K1A0R6 } 
 
\author{Sergei  Studenikin} 
\email[Corresponding author, email: ]{sergei.studenikin@nrc.ca} 
\affiliation{Emerging Technology Division,  
National Research Council, Ottawa, Canada, K1A0R6 } 
 
\author{Marek Korkusinski} 
\affiliation{Emerging Technology Division,  
National Research Council, Ottawa, Canada, K1A0R6 } 
  
\author{Louis Gaudreau} 
\affiliation{Emerging Technology Division,  
National Research Council, Ottawa, Canada, K1A0R6 } 
 
\author{Piotr Zawadzki} 
\affiliation{Emerging Technology Division,  
National Research Council, Ottawa, Canada, K1A0R6 } 
 
\author{Andy S. Sachrajda} 
\email[Corresponding author, email: ]{andy.sachrajda@nrc.ca} 
\affiliation{Emerging Technology Division,  
National Research Council, Ottawa, Canada, K1A0R6 } 
 
\author{Lisa Tracy} 
\affiliation{Sandia National Laboratories, Albuquerque, New Mexico 
  87185, USA } 
 
\author{John Reno} 
\affiliation{Center for Integrated Nanotechnologies,  
Sandia National Laboratories, Albuquerque, New Mexico 87185, USA}  
 
\author{Terry Hargett} 
\affiliation{Center for Integrated Nanotechnologies,  
Sandia National Laboratories, Albuquerque, New Mexico 87185, USA}  
 
\date{\today}


\begin{abstract}
We perform Landau-Zener-St\"uckelberg-Majorana (LZSM) spectroscopy on
a system with strong spin-orbit interaction (SOI), realized as a
single hole confined in a gated double quantum dot.
In analogy to the electron systems, at magnetic field $B=0$ and high
modulation frequencies we observe the photon-assisted tunneling (PAT)
between dots, which smoothly evolves into the typical LZSM
funnel-shaped interference pattern as the frequency is decreased.
In contrast to electrons, the SOI enables an additional, efficient
spin-flipping interdot tunneling channel, introducing a distinct
interference pattern at finite $B$.
Magneto-transport spectra at low-frequency LZSM driving show the two
channels to be equally coherent.
High-frequency LZSM driving reveals complex photon-assisted
tunneling pathways, both spin-conserving and spin-flipping,
which form closed loops at critical magnetic fields.
In one such loop an arbitrary hole spin state is inverted, opening the
way toward its all-electrical manipulation.
\end{abstract}
\maketitle

Currently there is interest in coherent control of
individual spins in the context of quantum-dot-based quantum
computing with spin qubits~\cite{nielsen,henneberger}. 
In gated devices, the single-spin control is achieved with
micromagnets~\cite{michel-micromagnets,laird,forster}  
or nanoantennas~\cite{koppens-nanoantenna,koppens-strip,dzurak},
requiring complex device engineering.
Alternative proposals envision rotating the spin of the moving carrier
via the electrostatically modulated Rashba spin-orbit interaction
(SOI)~\cite{datta-das}, which promises simpler designs and improved
device scalability. 
In electronic quantum dot systems this electrical spin manipulation 
has been reported in gated GaAs~\cite{schreiber,nowack} and InAs
devices~\cite{nadj_nature,pfund}. 
However, here the SOI magnitude is small, comparable to the
strength of nuclear hyperfine interactions, but much smaller than the
interdot tunneling 
coupling~\cite{yacoby_so-hyp,kouwen_so-hyp,tarucha_so-hyp,ensslin-onee}.  
On the other hand, theoretical proposals involving
holes~\cite{peeters_prl,szumniak_prb}, motivated by
predictions of suppressed hyperfine interactions with nuclear 
spins~\cite{coish_prl,coish_prb,burkard_nature,loss_prl,tartakovskii,heiss_longt1,gerardot_longt1,brunner_longt1,fras,varwig,degreve},
are at very early stages of implementation.
In silicon-based hole devices, both
lateral~\cite{hamilton_si,bohuslavskyi} and nanowire
systems~\cite{zwanenburg_nanow,higgin_nanow}, 
control of the hole spin via the electric dipole spin resonance and
the spin Rabi oscillations have been demonstrated~\cite{voisin,maurand}. 
In GaAs-based devices the  anisotropies of the hole
tunneling current in a magnetic field~\cite{hamilton_gaas}
and the zero in-plane g-factor~\cite{bogan} were traced to the
presence of strong SOI. 
Signatures of a high rate of spin-flipping tunneling between the dots in a
double dot have been detected in magneto-transport~\cite{bogan},
but the consequences of strong SOI have not yet been studied at the
single-hole level. 

Here we use Landau-Zener-St\"uckelberg-Majorana (LZSM)
interferometry to probe the dynamics of a single hole confined in a
lateral double-dot device. 
This experimental technique has been used to study the coherent
phenomena in a variety of physical systems, from atomic to
superconducting~\cite{shevchenko_theory}.
In gated dots, LZSM involves applying microwave modulation to the
detuning between the dots and measure the resulting tunnel current
or charge configuration.
It has been applied to study the dynamics and
characterize quantitatively the coherence of electronic charge
qubits~\cite{kouwen_pat,schreiber_pat,zalba_pat,stehlik_pat,forster_pat,petta_pat,braakman_pat}.
We demonstrate that for a single hole at zero magnetic field
the LZSM phenomena arise from a single tunneling channel, analogous to
that of a single-electron system.
We recover all interference features seen in electronic samples.
In particular, we observe the smooth evolution of the
photon-assisted tunneling (PAT) pattern at high driving frequencies to
the characteristic funnel-shaped fringes at low frequencies.
A dramatic change in the spectra is seen in a nonzero magnetic field,
where at all driving frequencies we report the coexistence of
interference features generated by two tunneling channels, one
spin-conserving and one spin-flipping. 
High-frequency LZSM interferometry reveals that
several microwave-assisted tunneling pathways may coexist
for critical values of detuning and magnetic fields,
offering novel regimes of control of the hybrid charge-spin system.

\begin{figure}[t]
\subfloat{ \includegraphics[width=0.22\textwidth]{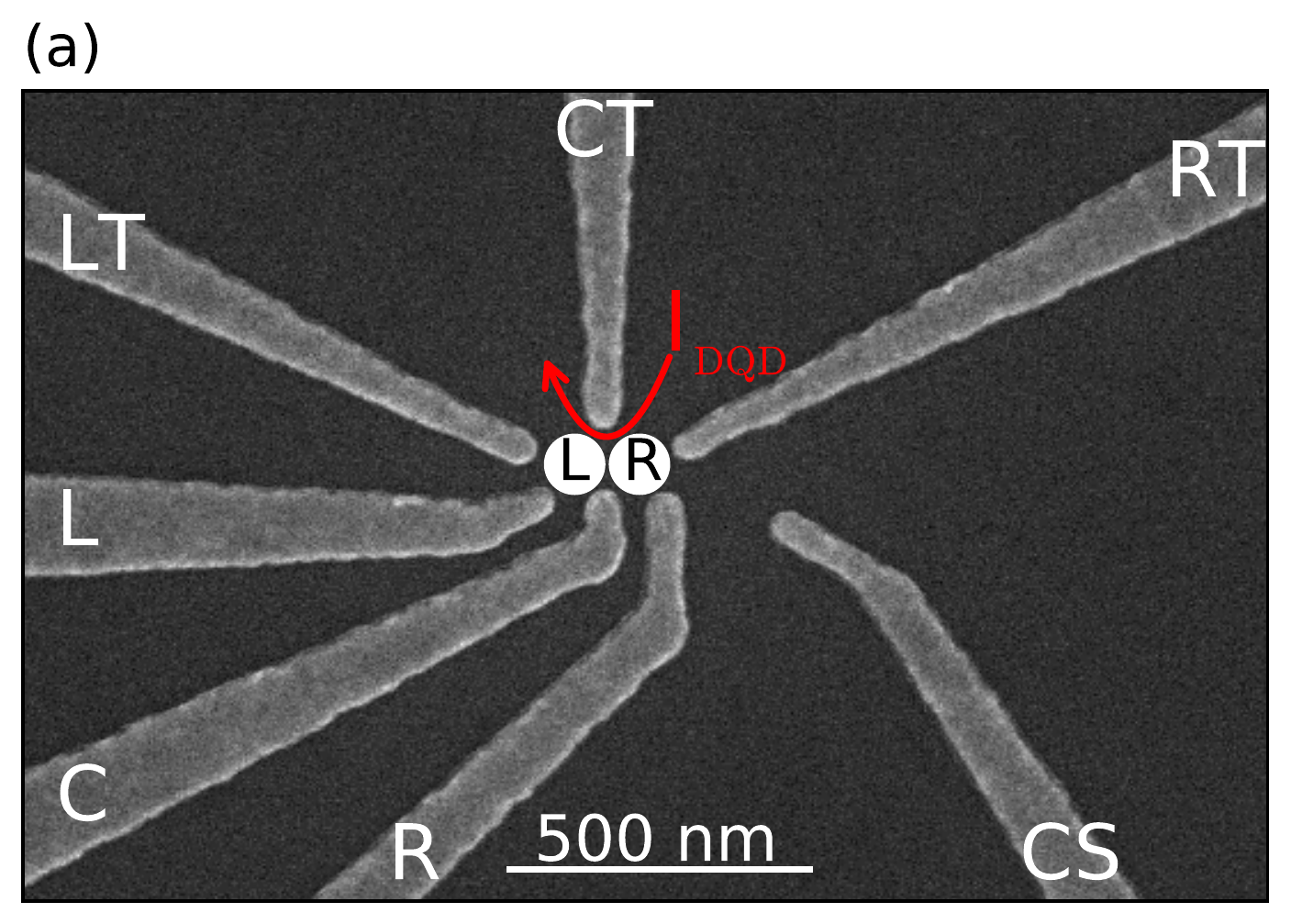} }
\hfill
\subfloat{ \includegraphics[width=0.22\textwidth]{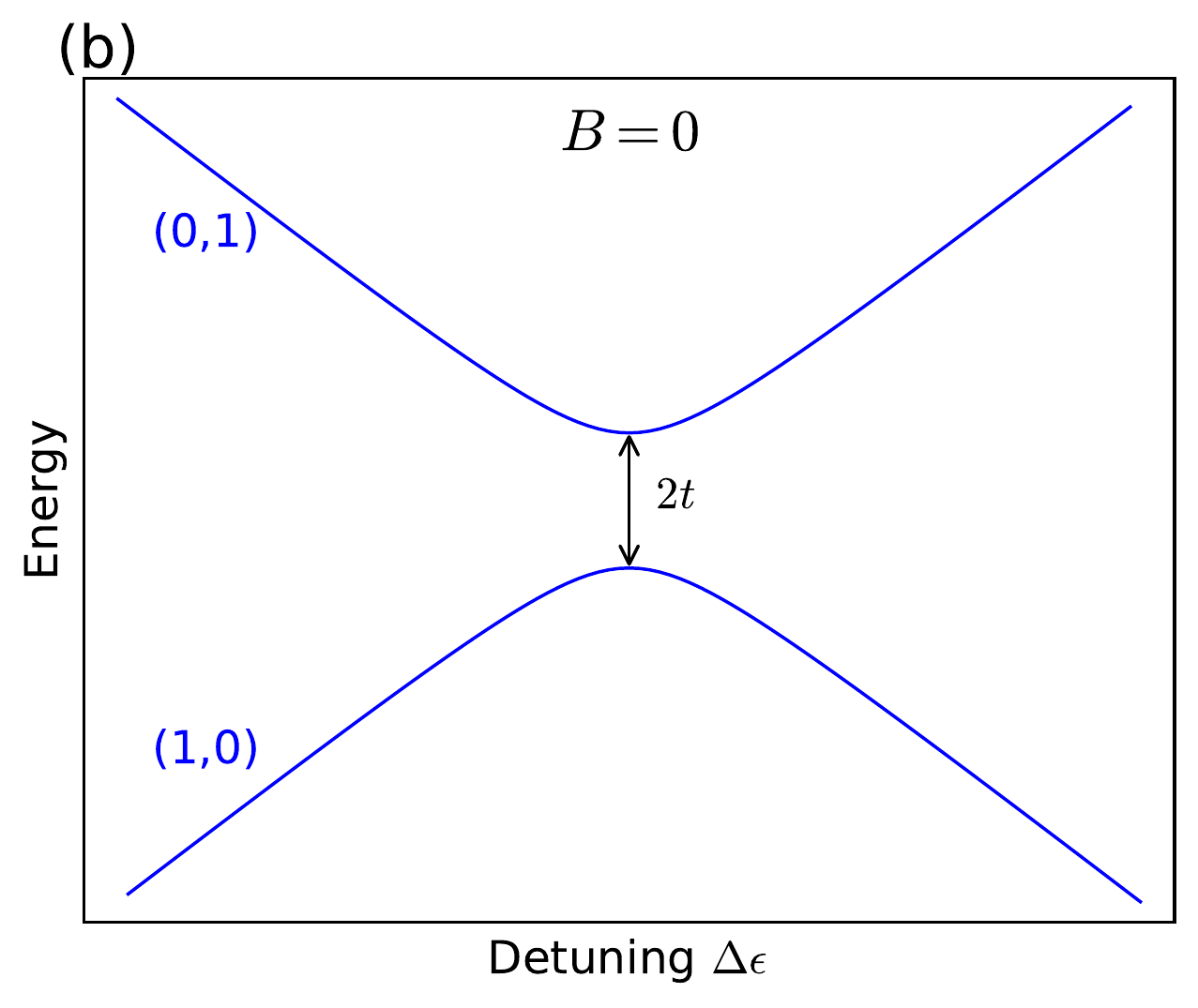} }
\\
\subfloat{ \includegraphics[width=0.22\textwidth]{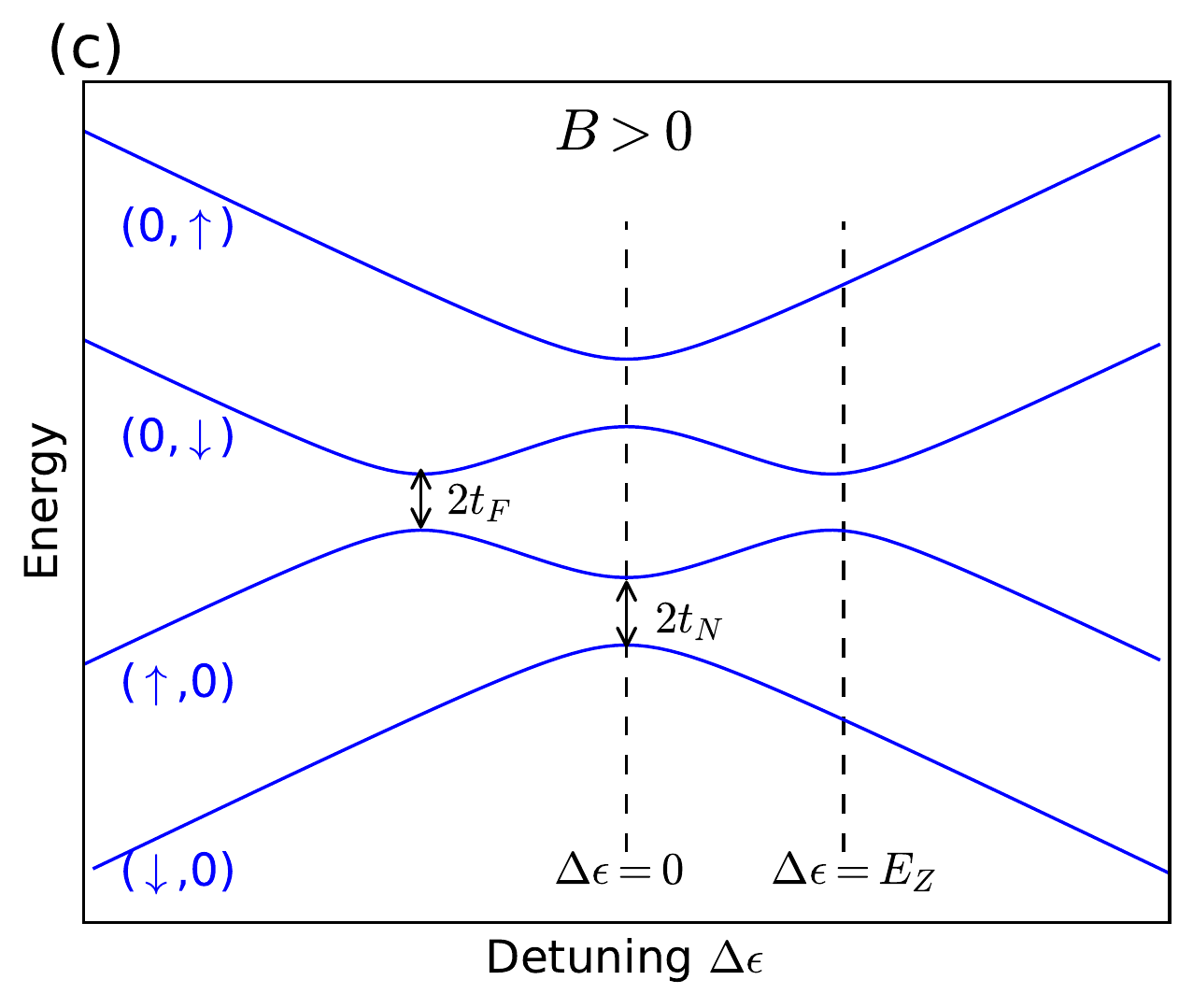} }
\hfill
\subfloat{ \includegraphics[width=0.22\textwidth]{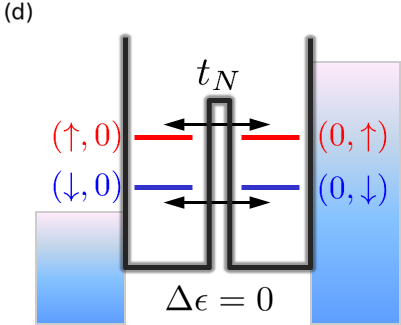} }
\\
\subfloat{ \includegraphics[width=0.22\textwidth]{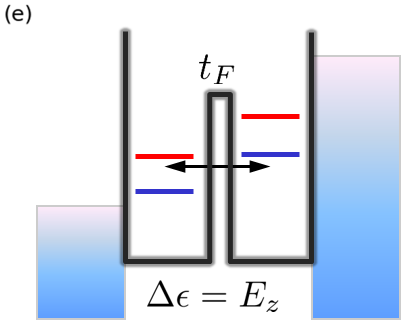} }
\hfill
\subfloat{ \includegraphics[width=0.22\textwidth]{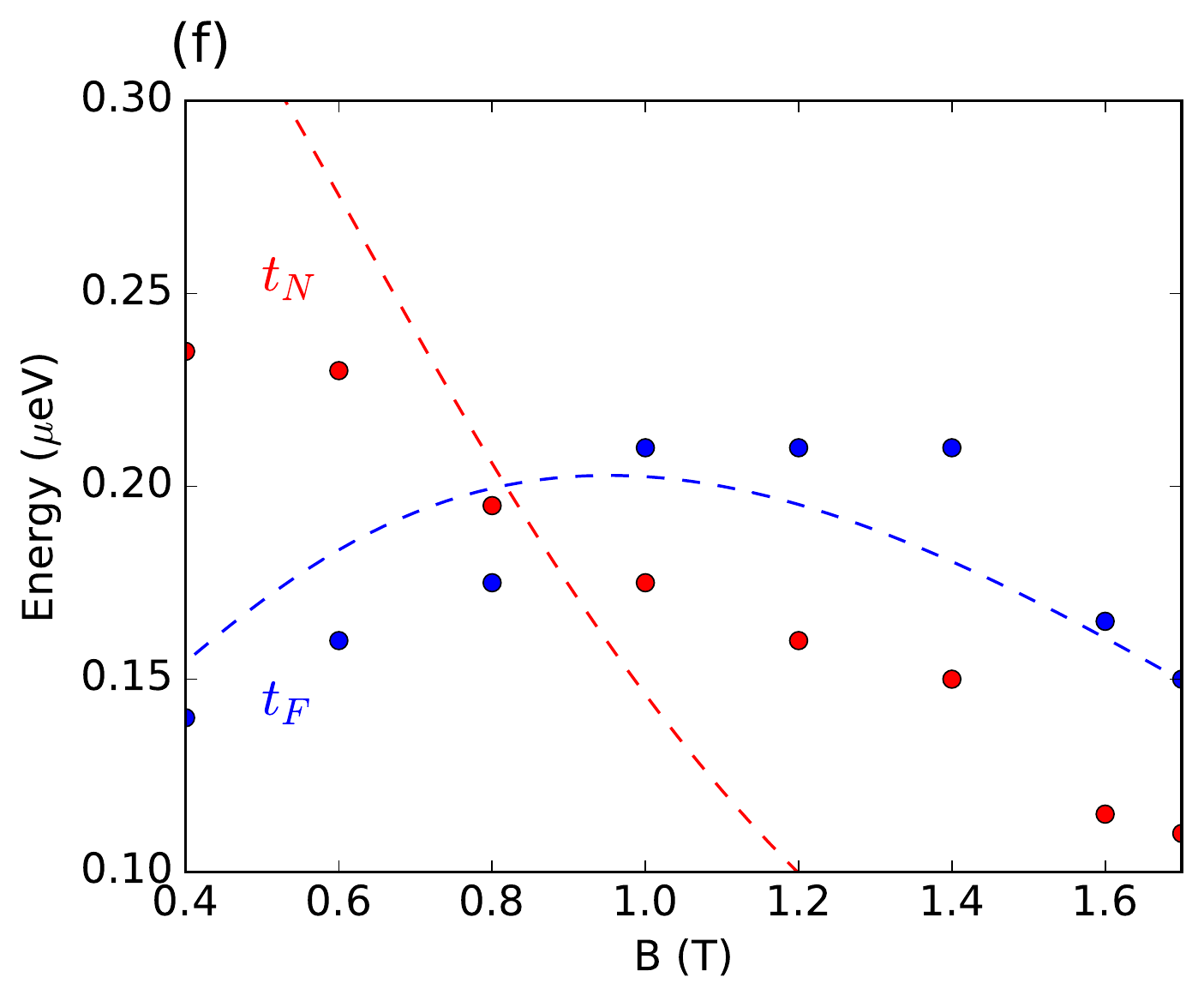} }
\caption{(a) Layout of the gates creating the double-dot lateral
  confinement.
The red arrow indicates the flow of the current $I_{DOT}$ in the
transport measurement. 
The yellow arrow denotes the current $I_{CS}$ through the charge
sensor $CS$.
Panels (b) and (c) show the energy diagram of the system as a function
of the detuning at zero and nonzero magnetic field, respectively.
Panels (d) and  (e) show schematic energy diagrams at a nonzero
magnetic field, for the detuning euqal zero and the Zeeman energh
$E_Z$, respectively. Matrix elements $t_N$ ($t_F$) characterize the
spin conserivng (flipping) tunneling resonance.
(f)  Magnitudes of the tunneling elements $t_N$ (black) and $t_F$
(red) as a function of the magnetic field extracted from experiment
(symbols) and predicted by microscopic theory (dashed lines).}
\label{fig1}
\end{figure}

Figure~\ref{fig1}(a) shows the gate layout of our GaAs lateral double
dot~\cite{bogan,tracy_APL}. 
We confine precisely one hole in the resultant lateral confinement,
and control electrically the charge state from $(n_L,n_R)=(1,0)$
to $(0,1)$, where $n_{L(R)}$ is the number of holes in the left-hand
(right-hand) dot.
In what follows we assume that the hole can be placed in each dot 
in two spin states, which results in four possible spin and charge
configurations: $(\downarrow,0)$, $(\uparrow,0)$, $(0,\downarrow)$,
and $(0,\uparrow)$.
At zero magnetic field the two spinors in each dot are degenerate.
Denoting the energies of $(1,0)$ and $(0,1)$ charge configurations
by $\varepsilon_L$ and $\varepsilon_R$, respectively,
the relative detuning $\Delta\varepsilon=\varepsilon_L-\varepsilon_R$
can be tuned by adjusting the voltage on the gate R.
Figure~\ref{fig1}(b) shows the energy diagram of the system as a
function of $\Delta\varepsilon$.
It consists of two degenerate energy levels, which anticross close to
resonance, i.e., when $\Delta\varepsilon$ approaches zero, with the
gap defined by the interdot tunneling rate $t$.
This diagram is identical to that for double-dot with a single
electron~\cite{kouwen_pat,schreiber_pat,zalba_pat,stehlik_pat,forster_pat,petta_pat,braakman_pat}.
The dramatic differences can be seen at a finite magnetic field
(Fig.~\ref{fig1}(c)).
Here, the different spin states of the same charge configuration are
separated by the Zeeman energy $E_Z$.
The states with the same spin exhibit anticrossings at
$\Delta\varepsilon=0$, as is the case for the electron, with the gaps
defined by the spin-conserving tunneling rate $t_N$.
The alignment of dot levels corresponding to that resonance condition
is visualized in Fig.~\ref{fig1}(d), in which the spin-down (up)
levels are represented with blue (red) bars.
In our hole system, the strong SOI enables two additional large
anticrossings occurring between the states with opposite spin, with
gaps defined by the spin-flipping tunneling rate $t_F$.
This element is due to the
SOI~\cite{peeters_prl,szumniak_prb,bulaev_loss_prl,bulaev_loss_prl2}
and its microscopic form is established in Ref.~\cite{supplement}. 
The alignment of levels occurring at the detuning $\Delta\varepsilon =
E_Z$ is shown schematically in Fig.~\ref{fig1}(e).
Here the left-dot spin-up level $(\uparrow,0)$ is resonant with
the right-dot spin-down level $(0,\downarrow)$ and we observe the
coherent spin-flipping tunneling resonance.

Following Refs.~\cite{hamilton_gaas,bulaev_loss_prl,bulaev_loss_prl2},
we describe our four-level system with the perturbative heavy-hole
Hamiltonian, which in the presence of the magnetic field perpendicular
to the dot surface is 
\begin{equation}
\hat{H} = \left[
\begin{array}{ccccc}
\varepsilon_L + E_Z / 2 & 0 & -t_N & -i t_F  \\
0 & \varepsilon_L - E_Z/2 & -i t_F  & -t_N \\
-t_N & i t_F  & \varepsilon_R +E_Z / 2 & 0 \\
i t_F  & -t_N & 0 & \varepsilon_R - E_Z / 2 \\
\end{array}
\right].
\label{hamil}
\end{equation}
The Zeeman energy $E_z = g^* \mu_B B$, where $g^*$ is the
effective hole g-factor, $\mu_B$ is the Bohr magneton, and $B$ is the
magnetic field.
We have estimated the numerical values of the elements of the above
Hamiltonian in magneto-transport spectroscopy by measuring the
tunneling current, depicted in Fig.~\ref{fig1}(a) with the red arrow,
as a function of the detuning $\Delta\varepsilon$ and the magnetic
field. 
We chose high source-drain voltage, corresponding to the alignment of
Fermi energies of the leads as represented in Fig.~\ref{fig1}(d), (e)
by striped boxes. 
Following the procedure outlined in the supplementary material,
Ref.~\cite{supplement}, in Fig.~\ref{fig1}(f) we show the extracted
dependence of $t_N$ and $t_F$ on the magnetic field (red and black
dots, respectively).
We compare it qualitatively to the trends predicted by a simple
microscopic model (red and black dashed lines, respectively).
The spin-flipping tunneling element is of similar magnitude
to that of the spin-conserving process, but the two elements differ in
their dependence on the magnetic field.
The element $t_N$ decreases, while $t_F$ first increases and then
decreases as the field grows.  
The behavior of $t_N$ is a consequence of the decreasing overlap 
between the left-dot and right-dot orbitals due to the
tightening of the cyclotron orbits as the field grows.
The complex dependence of $t_F$, on the other hand, is a direct
consequence of the SOI nature of this element, as it depends on the
orbital overlap as well as the momentum.
As the field grows, the overlap between orbitals decreases, but
the momentum obtains  an increasing correction arising from the magnetic
vector potential.
The observed behaviour is a result of the interplay of these two
trends. 
The characteristic increase of the spin-flipping tunneling element has
been observed for electronic double dots~\cite{tarucha_so-hyp}.

\begin{figure}[t]
\subfloat{ \includegraphics[width=0.24\textwidth]{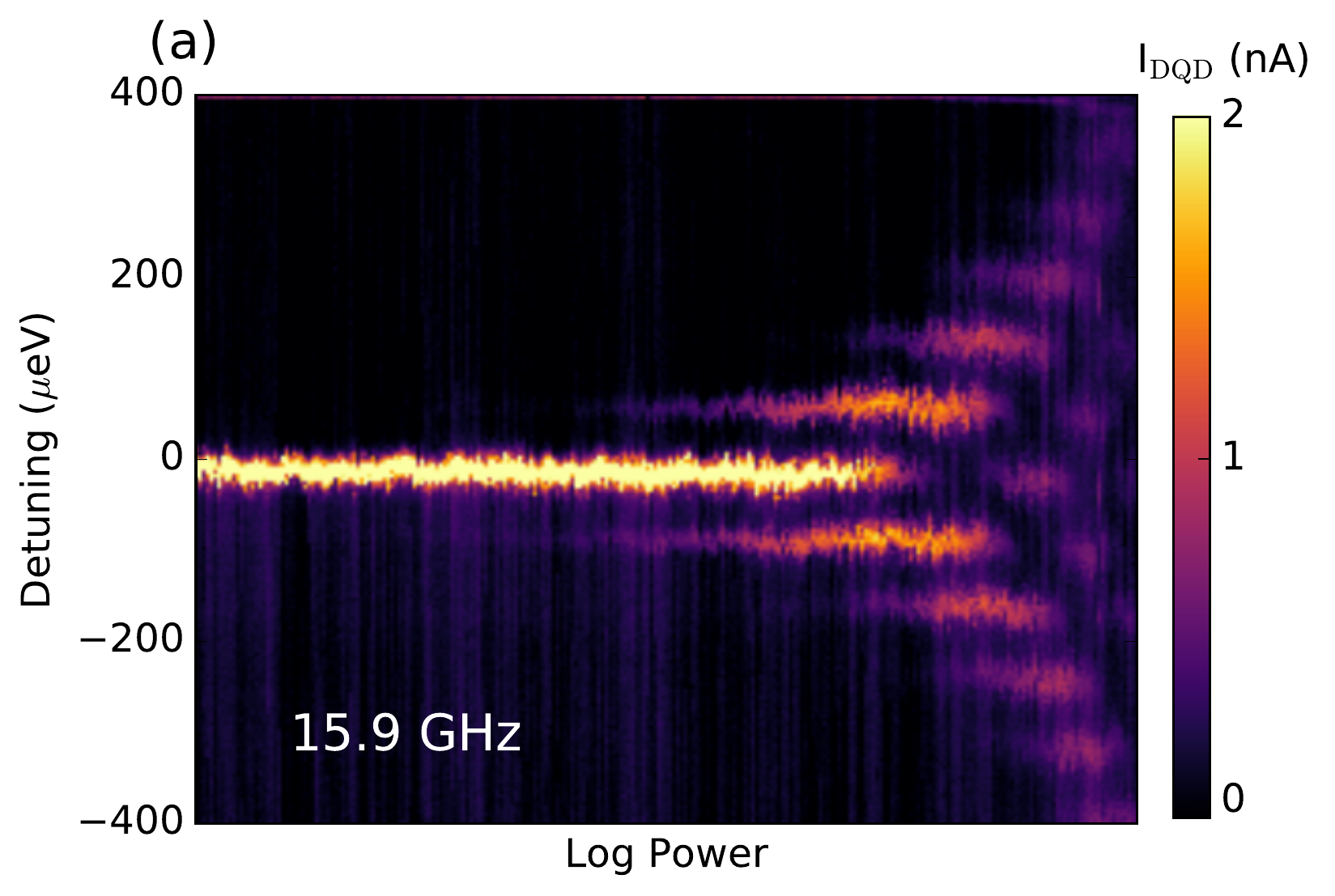} }
\hfill
\subfloat{ \includegraphics[width=0.205\textwidth]{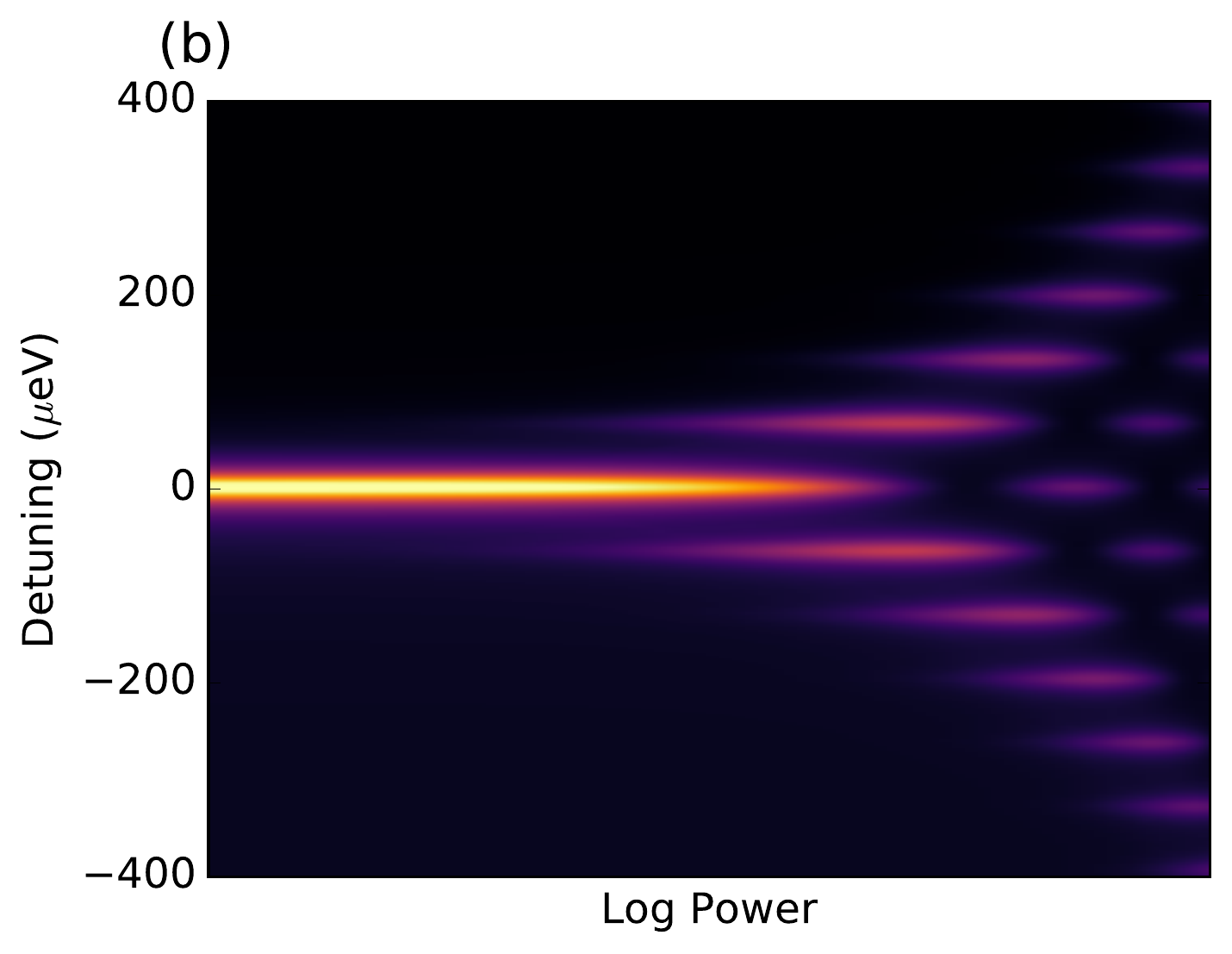} }
\\
\subfloat{ \includegraphics[width=0.24\textwidth]{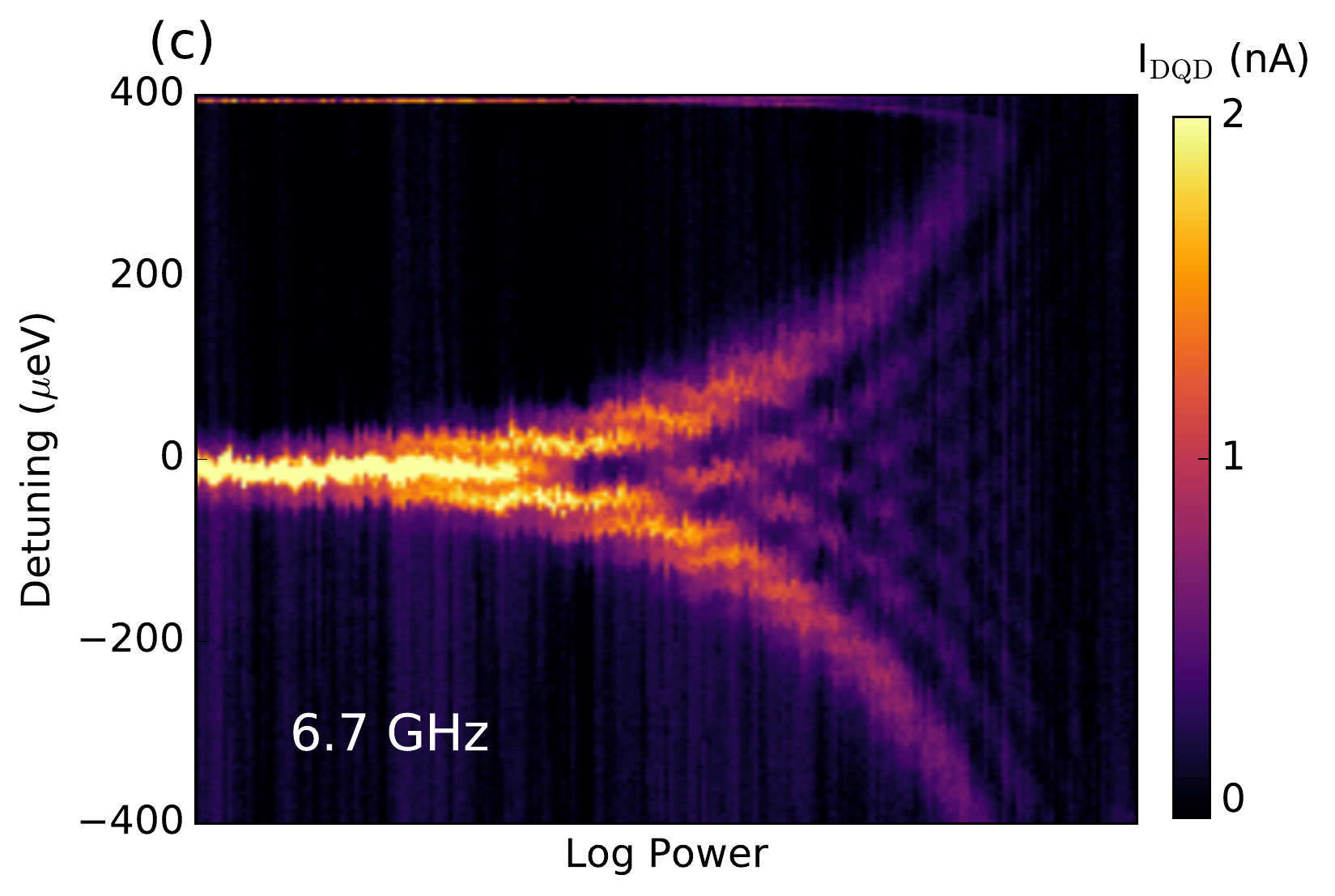} }
\hfill
\subfloat{ \includegraphics[width=0.205\textwidth]{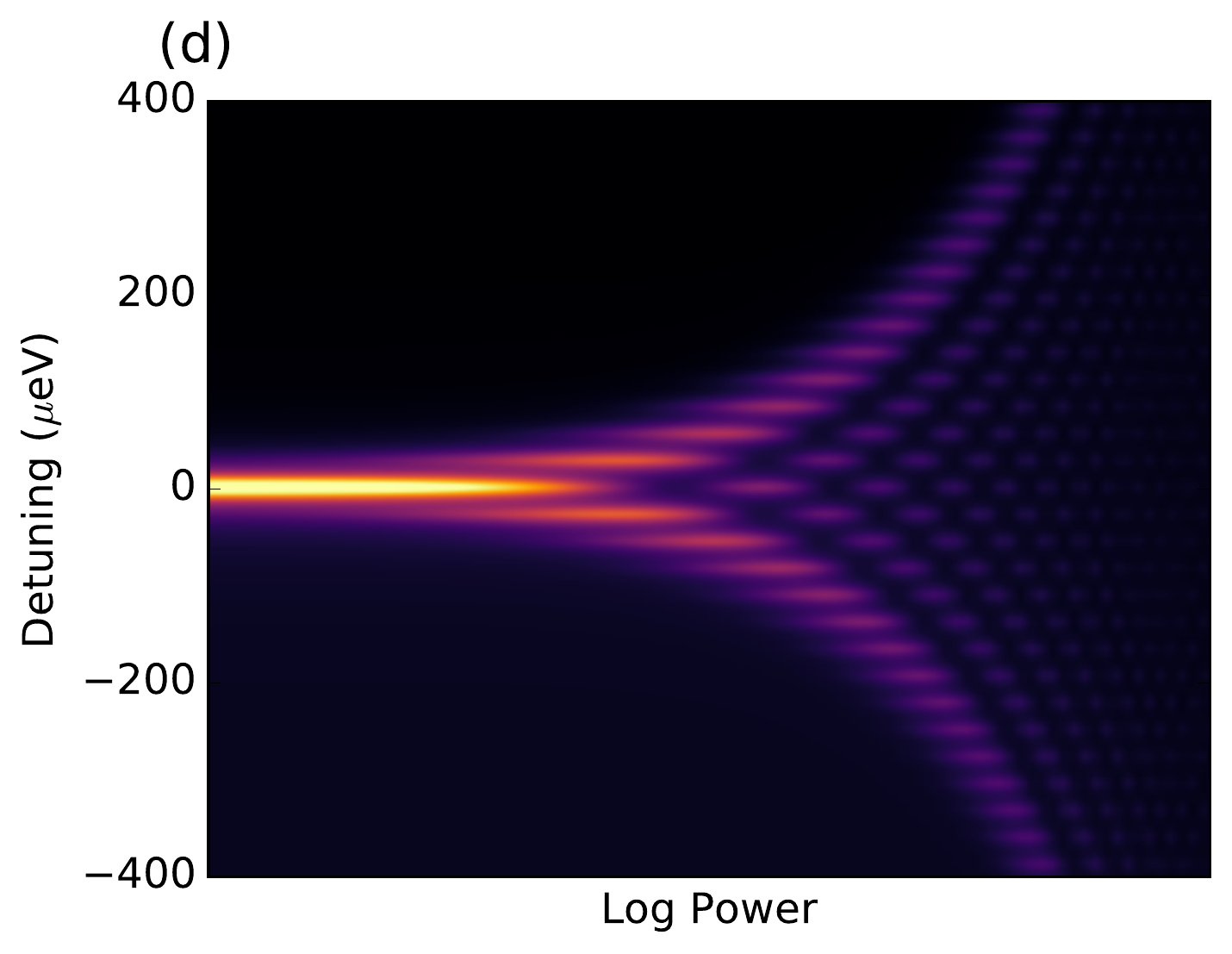} }
\\
\subfloat{ \includegraphics[width=0.24\textwidth]{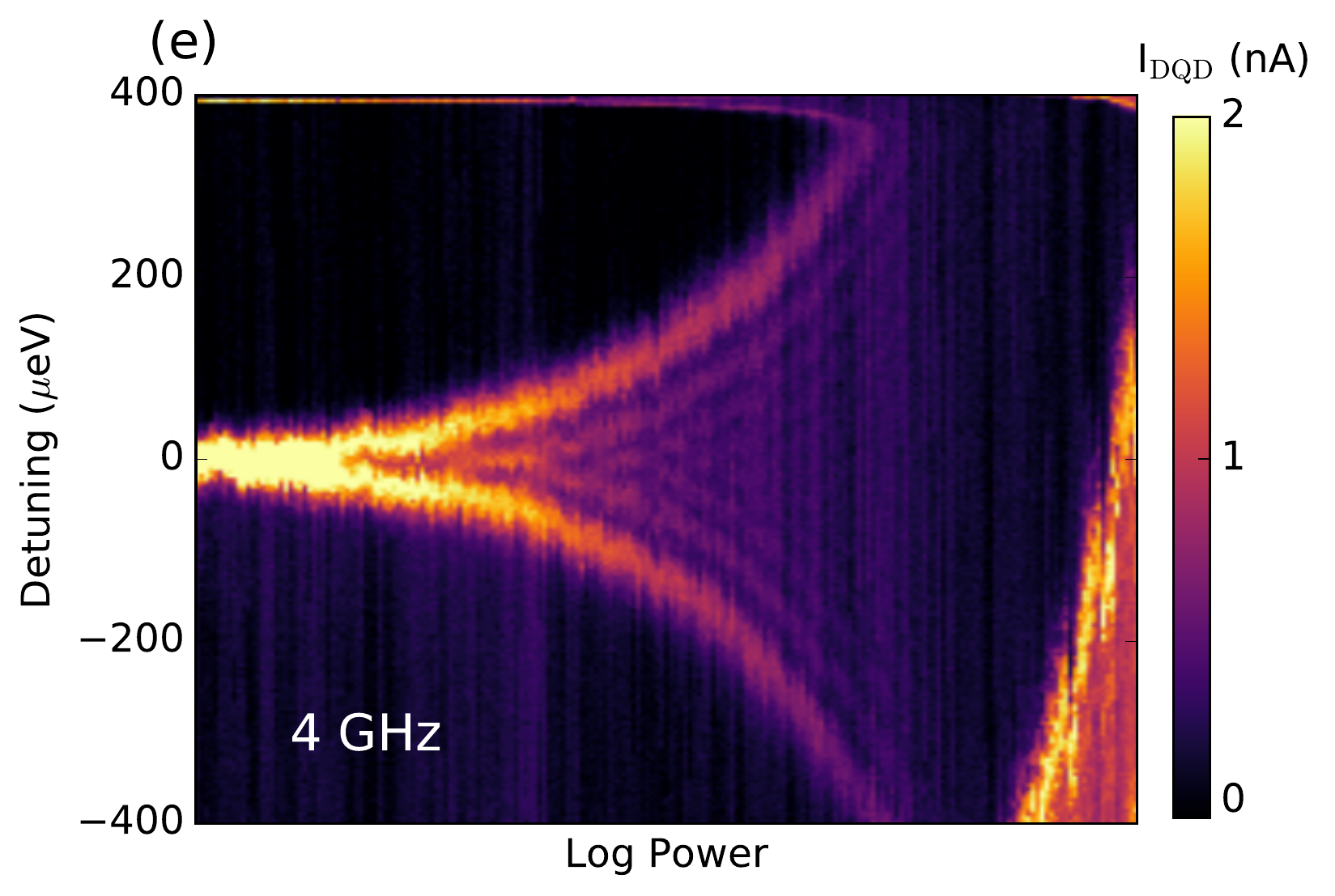} }
\hfill
\subfloat{ \includegraphics[width=0.205\textwidth]{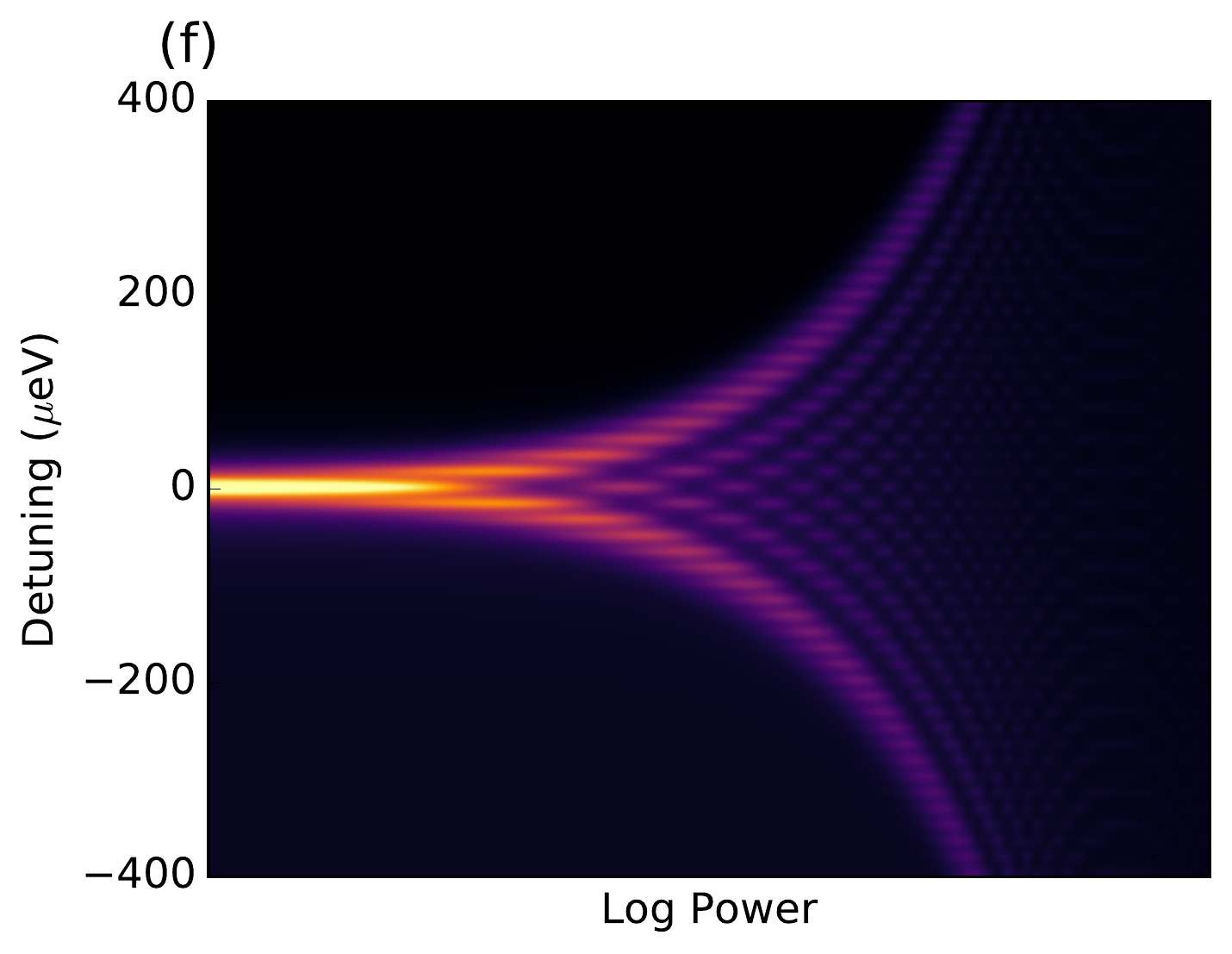} }
\caption{LZSM interferometry of a single hole at zero magnetic field.
Panels (a), (c), and (e) show the tunneling current measured as a
function of detuning and microwave power at driving frequencies $15.9$
GHz, $6.7$ GHz, and $4$ GHz, respectively.
Panels (b), (d), and (f) show respectively the results of model
calculations at matching conditions.}
\label{fig2}
\end{figure}

The LZSM interferometry in magneto-transport is performed
by applying a sinusoidal microwave modulation 
to the gate $R$ close to the right-hand dot.
We account for it by replacing the right-dot orbital energy
$\varepsilon_R \rightarrow \varepsilon_R + V_0 \sin(2\pi f t)$, 
where $V_0$ and $f$ are respectively the modulation
amplitude and frequency.
To make contact with earlier studies on electronic
double-dots~\cite{kouwen_pat,schreiber_pat,zalba_pat,stehlik_pat,forster_pat,petta_pat,braakman_pat},
we start at zero magnetic field.
Figure~\ref{fig2}(a) shows the tunneling current as a function of the
detuning $\Delta\varepsilon$ and microwave power for a high modulation
frequency $f=15.9$ GHz.
The set of interference fringes separated  in detuning by $2\pi\hbar f$
corresponds precisely to the PAT pattern studied in electronic
devices, with characteristic oscillations in intensity as a function
of power.
The pattern is reproduced theoretically in Fig.~\ref{fig2}(b) by
calculating the time-averaged current in the density-matrix rate equation
approach~\cite{supplement,shevchenko_theory,platero_theory,platero2}
applied to our four-level model.
As the microwave frequency is reduced to $6.7$ GHz, the PAT fringes
are closer in detuning as seen in experiment, Fig.~\ref{fig2}(c) and
theory, Fig.~\ref{fig2}(d).
The interference fringes are broadened by
decoherence~\cite{shevchenko_theory} allowing us to extract the value
of $T_2^*\approx 60$ and $75$ ps for $f=15.9$ GHz and $6.7$ GHz,
respectively.
For an even lower frequency $f=4$ GHz [Fig.~\ref{fig2}(e) experiment
and (f) theory] the fringes coalesce and form funnel-shaped features
related to the LZSM spectra recorded in electronic systems modulated
by single pulses~\cite{petta_lzs,studenikin_lzs,poulin_lzs,korkusinski_lzs}.
Here the extracted $T_2^*=90$ ps, the apparent systematic increase being
due most likely to decreased charge noise at lower microwave
frequencies. 

\begin{figure}
\subfloat{ \includegraphics[width=0.22\textwidth]{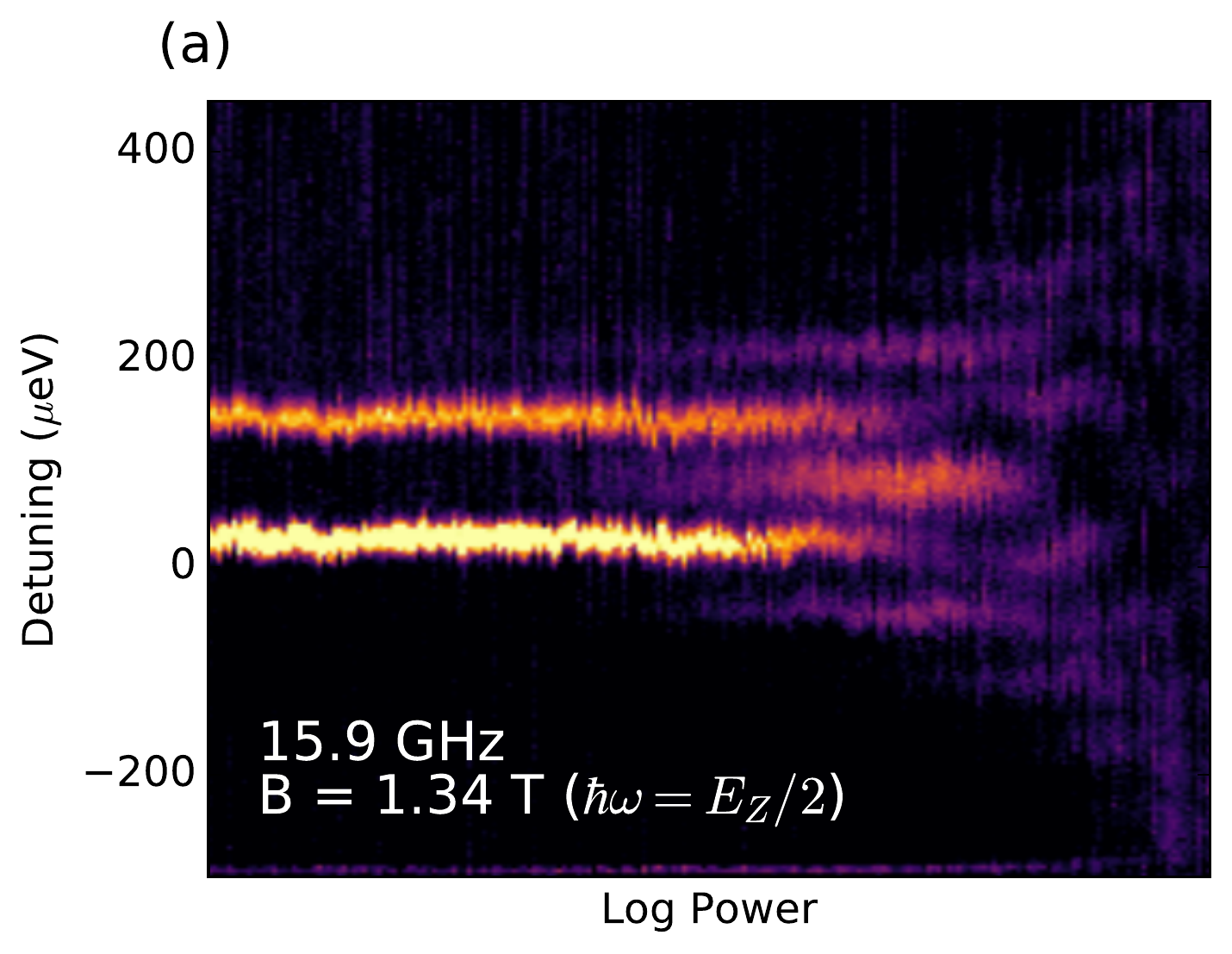} }
\hfill
\subfloat{ \includegraphics[width=0.22\textwidth]{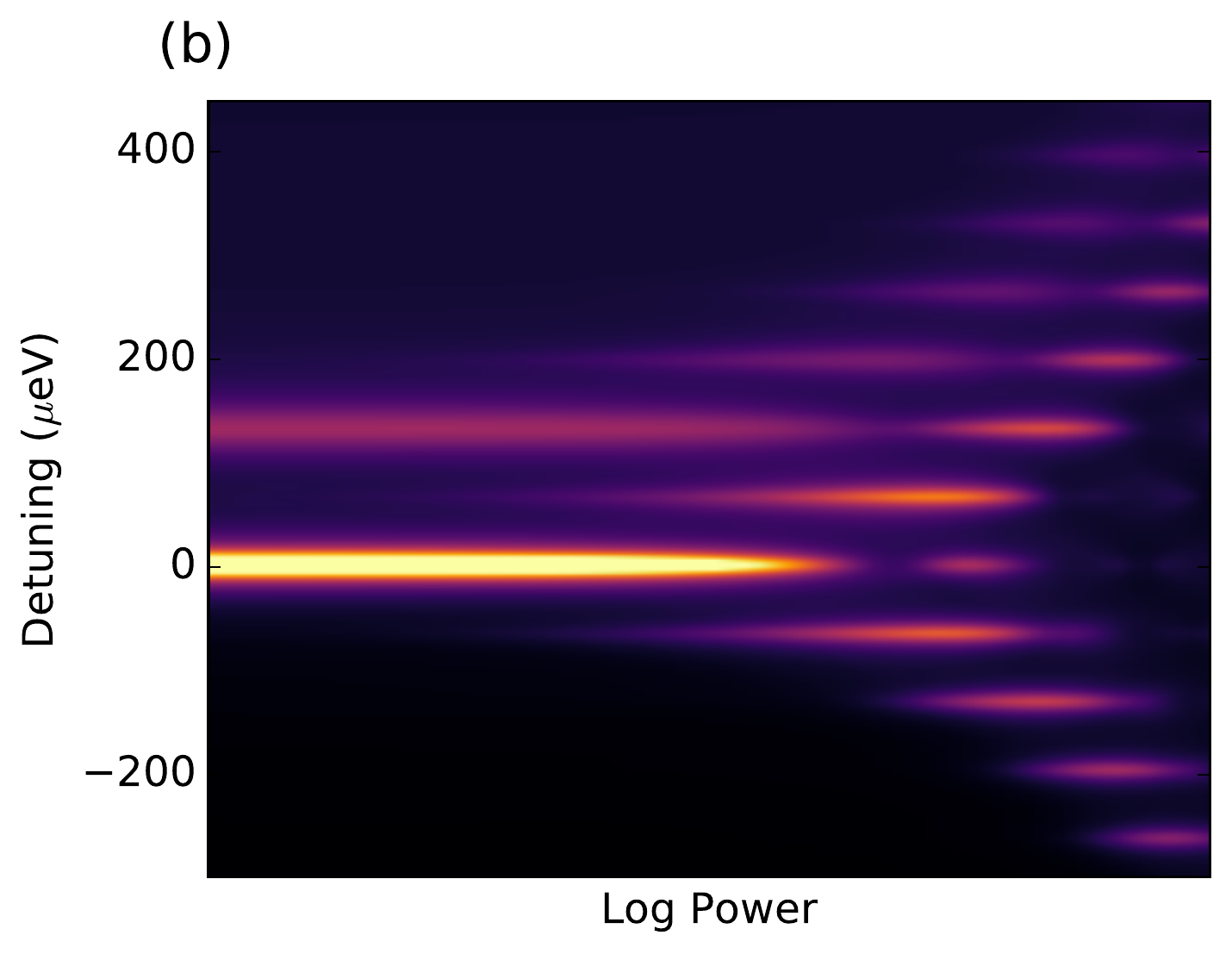} }
\\
\subfloat{ \includegraphics[width=0.22\textwidth]{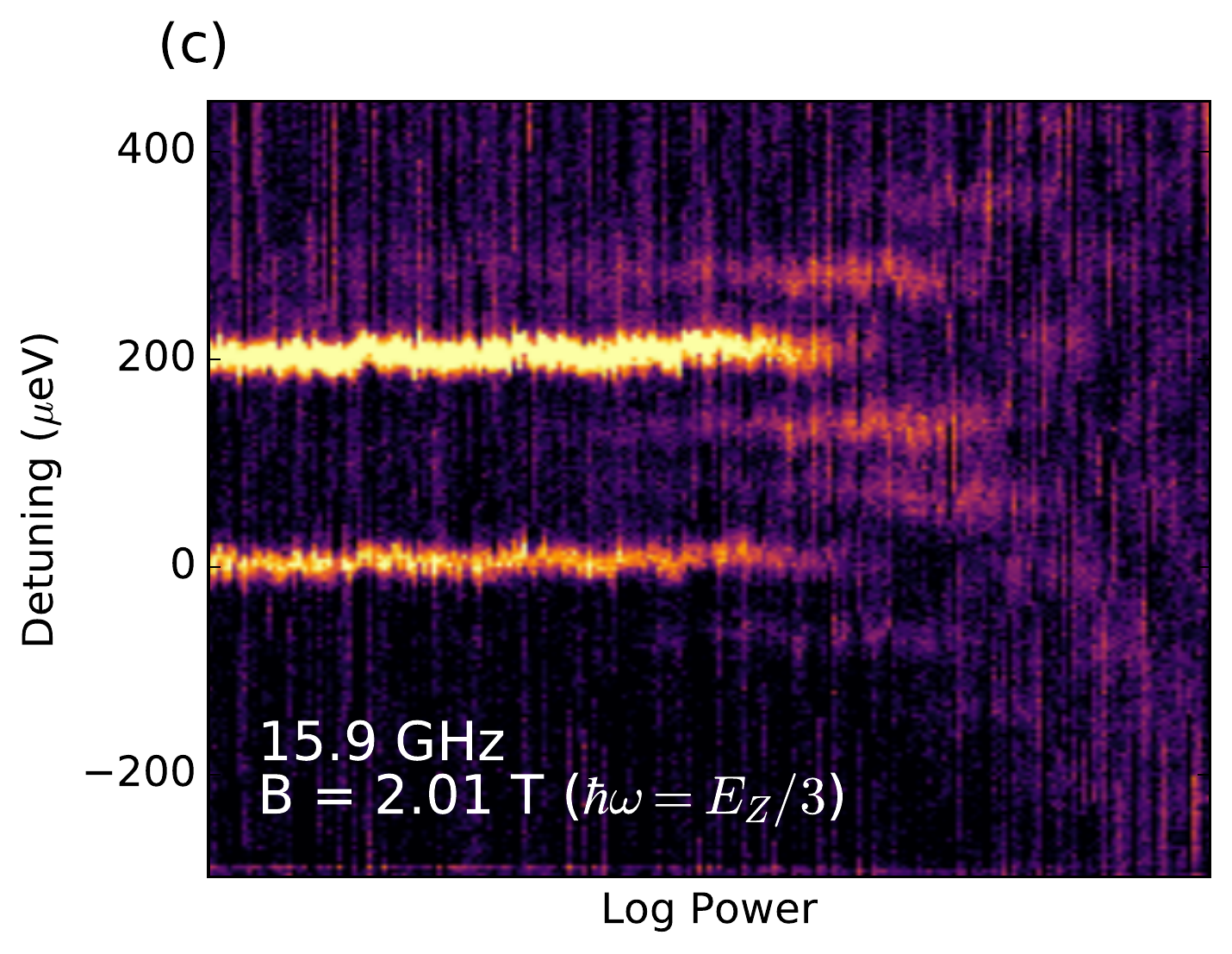} }
\hfill
\subfloat{ \includegraphics[width=0.22\textwidth]{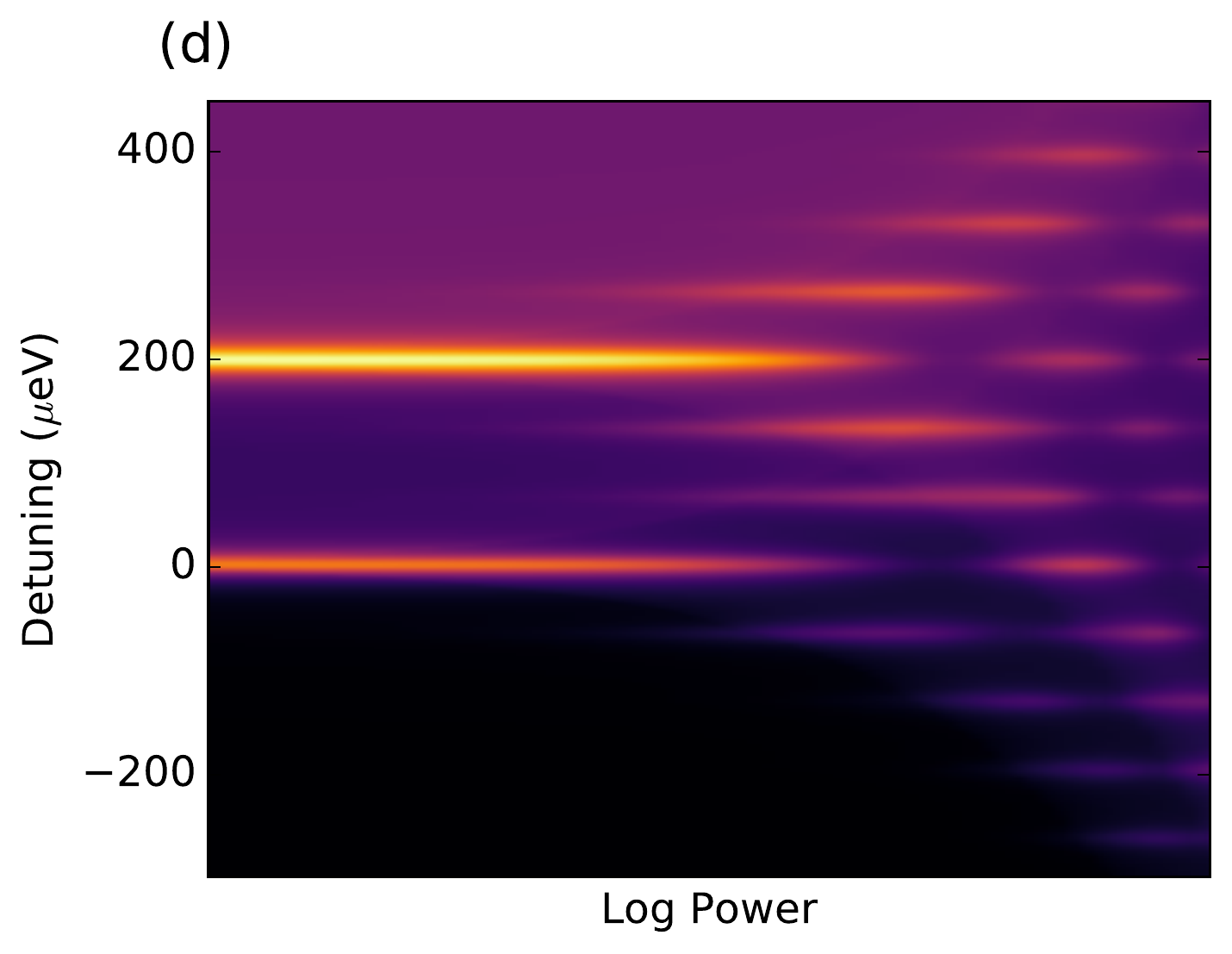} }
\\
\subfloat{ \includegraphics[width=0.22\textwidth]{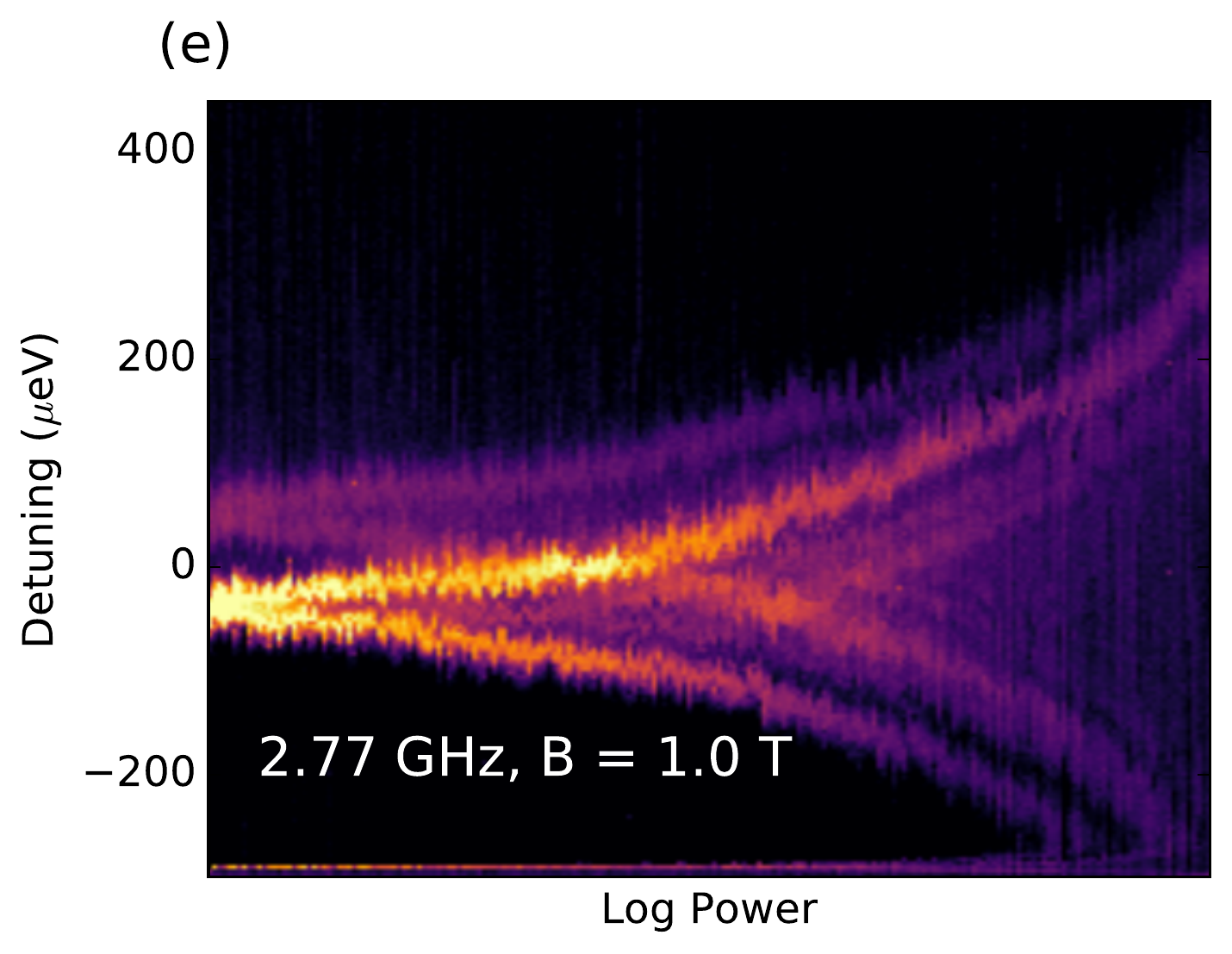} }
\hfill
\subfloat{ \includegraphics[width=0.23\textwidth]{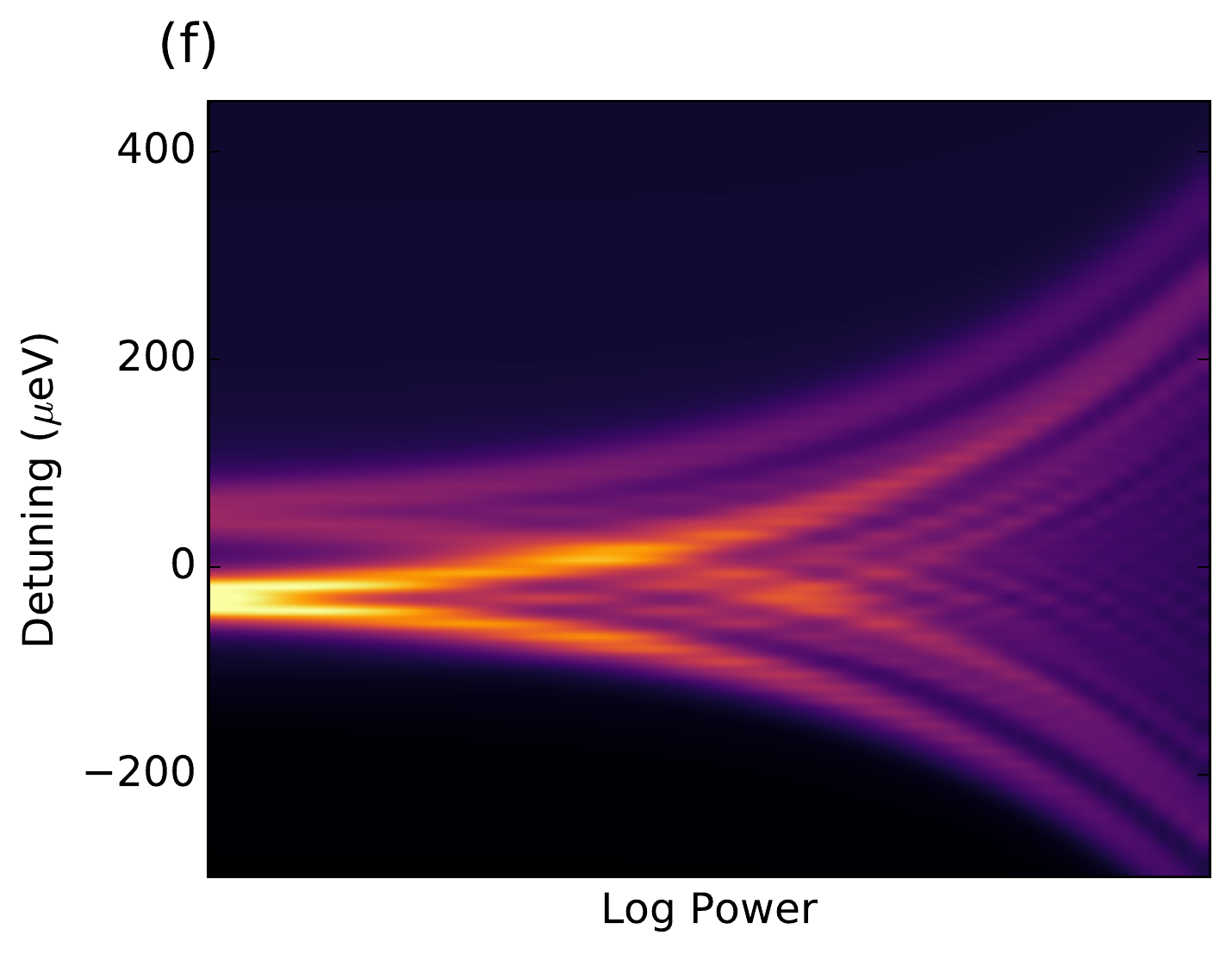} }
\caption{LZSM interferometry of a single hole at a nonzero magnetic
  field.
Tunneling current measured (a) and calculated (b) as a function of
detuning and microwave power for the driving frequency $f=15.9$ GHz
and the magnetic field $B=1.34$ T. The field is chosen so that the
Zeeman energy $E_Z=2hf$.
Panels (c) and (d) show the current at $B=2$ T, i.e., $E_Z=3hf$, with
the same microwave modulation. 
Panels (e) and (f) show respectively the tunneling current measured
and calculated at $B=1$ T for driving frequency $f=2.77$ GHz.}
\label{fig3}
\end{figure}

The LZSM spectra are dramatically different in a nonzero magnetic
field. 
Figures~\ref{fig3}(a) and (c) show respectively the tunneling current
as a function of microwave power for high frequency modulation
($f=15.9$ GHz) and two values of magnetic field, $B=1.34$ T and $B=2$
T, respectively.
Figure~\ref{fig3}(e) shows the current measured at $B=1$ T and a
low-frequency modulation ($f=2.77$ GHz).
Figures~\ref{fig3}(b), (d), and (f) show the results of corresponding
simulations.
Here the spin-conserving and spin-flipping tunneling channels become
nondegenerate resulting in the appearance of two sets of interference
fringes.
In Fig.~\ref{fig3}(a) we are in the PAT regime. 
Here the magnetic field was chosen so that the resulting Zeeman energy
$E_Z = 2 hf$, leading to the overlap of first-order interference
features of the two patterns. 
An increase of the magnetic field to $2$ T gives the Zeeman energy 
$E_Z = 3 hf$, resulting in a relative shift of these patterns in
detuning, as seen in Fig.~\ref{fig3}(c).
In Fig.~\ref{fig3}(e) we recover the characteristic low-frequency
funnel-shaped fringes at the field of $B=1$ T.
The current maxima evolve from the two-peak structure at low power,
through broader, but separate interference patterns at intermediate
powers, towards a complex, overlapping structure at high powers.
The LZSM driving introduces a precise timing metric, which allows to
fit the tunneling elements $t_N$ and $t_F$ as well as the
decoherence times $T_{2N}$ and $T_{2F}$ characterizing the
spin-conserving and spin-flipping tunneling process,
respectively~\cite{supplement}. 
We find nearly power-independent values of $t_N=0.26\pm 0.02$ $\mu$eV
and $t_F=0.28\pm 0.04$ $\mu$eV.  
Moreover, decoherence times $T_{2N}\approx T_{2F}=T_2^*$ appear to
decrease from $\sim 120$ ps down to $\sim 80$ ps as the power
increases, which is probably due to the heating of the sample.
The simulation result in Fig.~\ref{fig3}(d)  shows the current
calculated for a power-independent value of $T_2^*=90$ ps in excellent
agreement with experimental data. 
The existence of the LZSM interference pattern for both tunneling
processes indicates that the spin-flipping channel is as strong and as
coherent as the usual spin-conserving one.
As a result, we deal with a hybrid spin-charge qubit spanned in the
basis of four spin-orbital states, with the orthogonal degrees of
freedom being the position and the spin of the hole.
This system differs from the electronic singlet-triplet hybrid
qubit demonstrated recently in silicon~\cite{shi_eriksson}
in that here the tunneling of the hole from one dot to another can be
realized both without and with spin flip, offering enhanced
functionality and control.

\begin{figure}
\subfloat{ \includegraphics[width=0.22\textwidth]{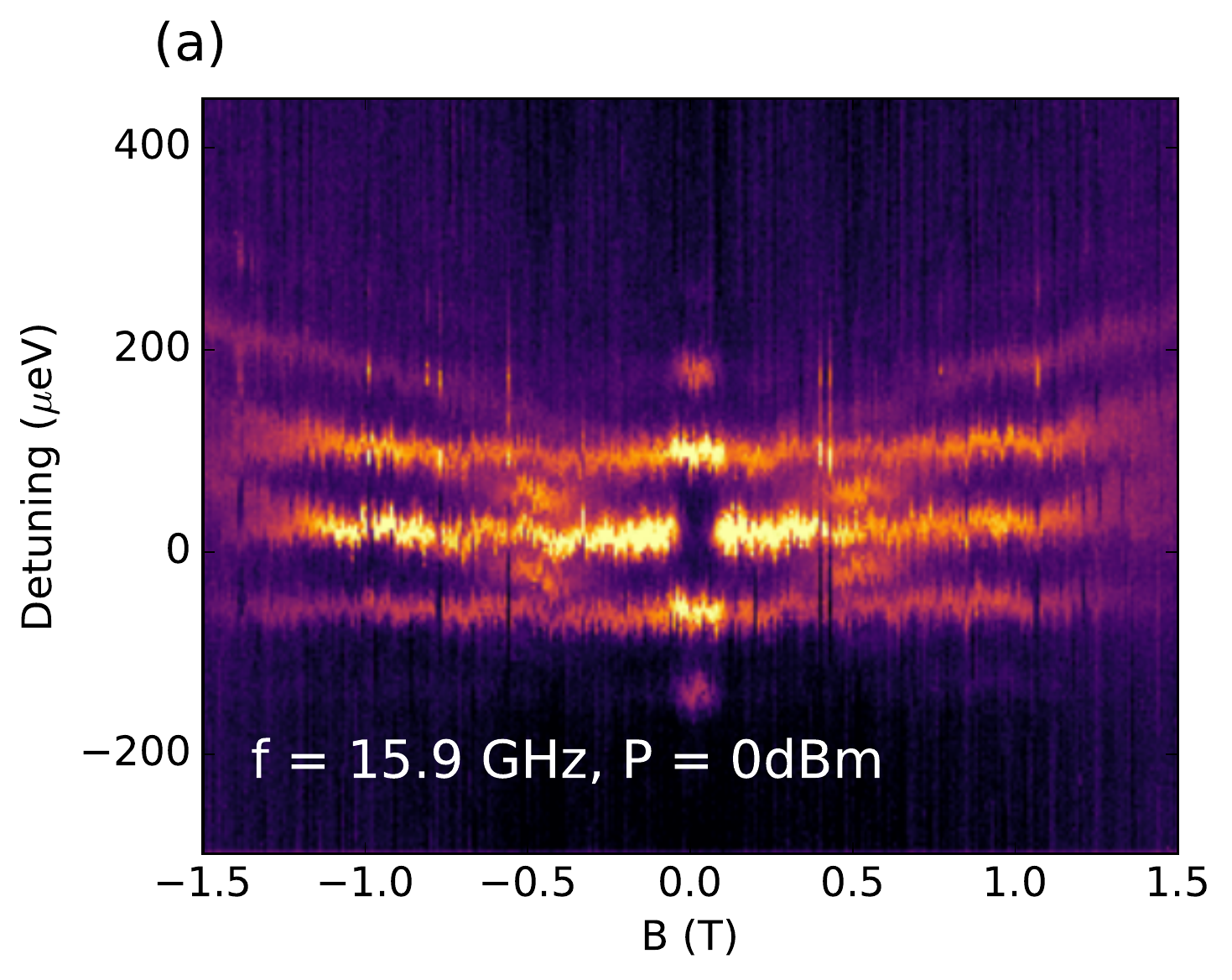} }
\hfill
\subfloat{ \includegraphics[width=0.22\textwidth]{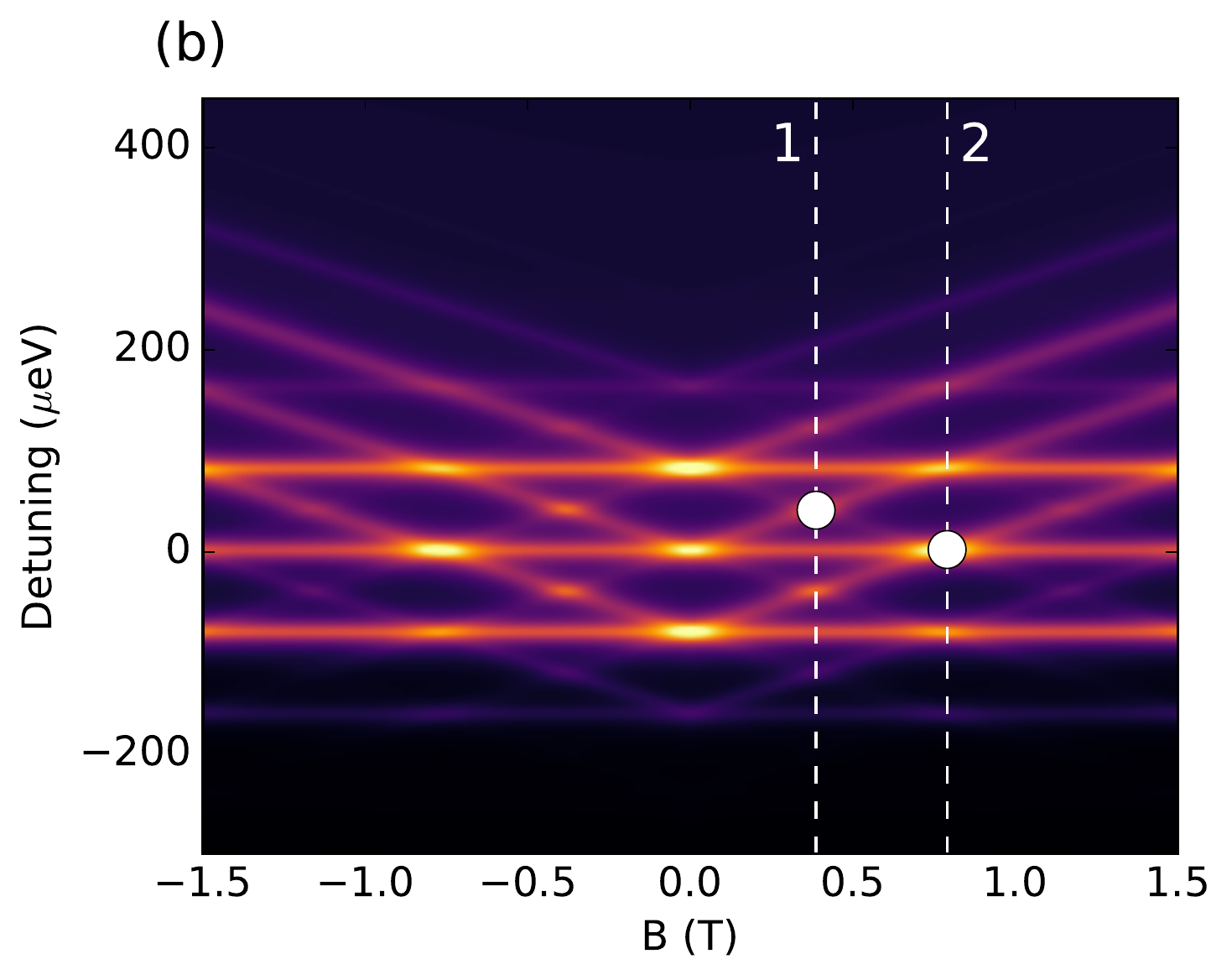} }
\\
\subfloat{ \includegraphics[width=0.22\textwidth]{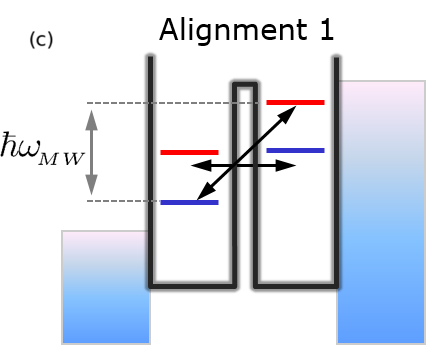} }
\hfill
\subfloat{ \includegraphics[width=0.22\textwidth]{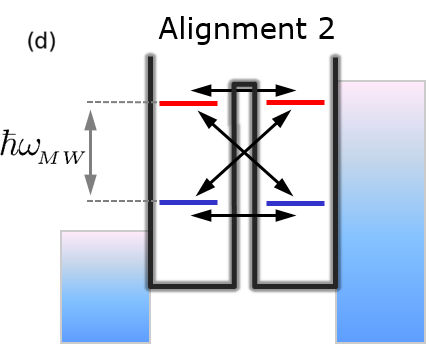} }
\caption{
(a), (b) Measured and calculated tunneling current as a function of
detuning (vertical axis) and magnetic field (horizontal axis) at
microwave frequency $f=19.56$ GHz and power $0$ dBm.
Panels (c) and (d) show the alignment of quantum dot levels and the
resulting tunneling pathways respectively at conditions (1) and (2)
from panel (b). }
\label{fig4}
\end{figure}

The relative shift of the two LZSM patterns can be tuned with the
magnetic field.
To demonstrate this, in Fig.~\ref{fig4}(a) we plot the PAT spectra at
the microwave power of $5$ dBm as a function of the magnetic field,
while panel (b) shows the result of simulation.
With an excellent agreement between the experiment and theory, we find
two clear families of lines.
The field-independent (horizontal) features correspond to the
spin-conserving transitions, while the field-dependent traces denote
the spin-flipping resonances.
The second family consists of lines with positive and negative slopes,
corresponding to the two spin-flip transitions possible when tunneling
from the right to the left dot, i.e., $(0,\uparrow) \rightarrow
(\downarrow,0)$ and $(0,\downarrow) \rightarrow (\uparrow,0)$.

Due to their different dependence on the magnetic field, the different
families exhibit intersections (degeneracies) at critical values of the
field and detuning.
This is the central result of our LZSM interferometry study.
In Fig.~\ref{fig4} we discuss the two possible types of these
intersections.
The first one, denoted in panel (b) as (1), involves two fringes, one
positively and one negatively sloped in the magnetic field.
The diagrammatic description of this condition is shown in panel (c).
Since the positions of both fringes are field-dependent, they both
correspond to spin-flipping tunneling, albeit at a different PAT
order (here, $n=0$ and $n=+1$).
The condition for its occurrence connects the detuning, the Zeeman
energy, and the microwave frequency as $\Delta\varepsilon = E_Z =
\pi\hbar f$.
Here, the two resonant tunneling channels are enabled simultaneously,
but remain independent, and the resultant tunneling current is a
simple sum of the two contributions.
This is why in Fig.~\ref{fig4}(a) and (b) we see a clear enhancement
(a bright spot) compared to the visibility of the linear features
leading into it.

The independence of the two channels makes it possible
to invert the arbitrary hole spin state
in tunneling from the right to the left dot.
Indeed, let us prepare the right-dot hole state in the form
$\alpha|(0,\uparrow)\rangle + \beta |(0,\downarrow)\rangle$.
At this resonance point, the state is transferred (within the LZSM
process) to the left-dot state 
$\beta|(\uparrow,0)\rangle + \alpha e^{i\phi} |(\downarrow,0)\rangle$,
with a possible phase factor $\phi$ dependent on the relative phase of
the spin-flipping and spin-conserving processes.
This enables a complete, electrically-controlled hole spin flip, 
while the control over the phase can be achieved e.g. through
engineering of the Rashba SOI.
In our LZSM study we apply a monochromatic microwave modulation,
causing the hole to tunnel back and forth between the dots.
However, the hole could be transferred to the left dot by employing
appropriate pulse shaping techniques.

The second type of intersections is denoted by the condition
(2) in Fig.~\ref{fig4}(b), and is visualized schematically in
panel (d).
At such three-fold intersections, spin-flipping and spin-conserving
channels of different interference orders are simultaneously active
In this case, $\Delta\varepsilon = 0$ and $E_Z = 2 \pi\hbar f$,
resulting in coincidence of $n=\pm 1$ spin-flipping and $n=0$
spin-conserving processes.
The three channels depcited in the diagram (d) share common spinors,
i.e., the process can be seen as a closed loop, evident from the
alignment of arrows in the diagram.
This opens a possibility of quantum interference, whereby the
tunneling current could be suppressed completely, even
though the energy matching conditions are met, in analogy to the
coherent population trapping process.
Theory predicts that the required balance of effective tunneling
matrix elements could be achieved by an appropriate choice of the
microwave power~\cite{supplement}.
We note that a demonstration of such interference would be challenging
in an electron system due to the disparity of the spin-conserving and
spin-flipping elements.

In summary, we have applied the LZSM interferometry to study the
coherent tunneling of a single hole between the dots of a gated
lateral double-dot device.
In a transport experiment at zero magnetic field we reproduced in all
detail the interference patterns known from the electronic systems,
both at high and low driving frequencies.
At finite fields we demonstrated the coexistence of two, equally
strong and equally coherent tunneling channels - one spin-conserving
and one spin-flipping, the latter enabled by the strong SOI in our
system. 
The resulting formation and coincidence of different photon-assisted
tunneling pathways was studied in high-driving frequency LZSM
interferometry.
One such coincidence enables a complete inversion (flip) of an
arbitrary spin state of the hole as it tunnels between the dots.
At the other complex resonance point the hole visits the spin
up and down orbitals of the left and right dots in a closed loop.
Such a closed transport pathway is a prerequisite for single-hole
quantum interference phenomena in analogy with the Aharonov-Bohm
effect.


Acknowledgment.
AB and SS thanks the Natural Sciences and Engineering Research Council
of Canada for financial support.  
This work was performed in part at the Center for Integrated
Nanotechnologies, a U.S. DOE, Office of Basic Energy Sciences user
facility, and Sandia National Laboratories, a multi-mission laboratory
managed and operated by National Technology and Engineering Solutions
of Sandia, LLC., a wholly owned subsidiary of Honeywell International,
Inc., for the U.S. Department of Energy's National Nuclear Security
Administration under contract DE-NA-0003525.


\pagebreak
\widetext
\begin{center}
\textbf{\large Supplementary material for  
Landau-Zener-St\"uckelberg-Majorana interferometry
of a single hole}
\end{center}
\setcounter{equation}{0}
\setcounter{figure}{0}
\setcounter{table}{0}
\setcounter{page}{1}
\makeatletter
\renewcommand{\theequation}{S\arabic{equation}}
\renewcommand{\thefigure}{S\arabic{figure}}
\renewcommand{\bibnumfmt}[1]{[S#1]}
\renewcommand{\citenumfont}[1]{S#1}

\section{Sample fabrication}

The experimental study was performed on a double quantum dot (DQD)
fabricated from an undoped GaAs/Al$_x$Ga$_{1-x}$As ($x$=50\%)
heterostructure (VA0670) employing lateral split-gate technology
[\onlinecite{sKomijani,sTracy-rep,sTracy,sbogan}].    
A suitable DQD potential profile was defined by the deposited lateral
Ti/Au gates.   
A scanning electron micrograph (SEM) of the gate layout is shown in
Fig. 1(a) of the main text.
Holes were generated by a global gate deposited above the structure
(not shown in Fig. 1(a) of the main text) separated by a 110 nm-thick
Al$_2$O$_3$ dielectric layer grown by an atomic layer deposition
technique.   
Left and right plunger gates, labeled as L and R, respectively, were
used to tune individually the hole potentials in each dot, while the  
central gate C was used to adjust the interdot tunneling barrier.   
The sample was cooled down in a dilution refrigerator at the  
nominal electron temperature $\sim$100 mK.  
For transport measurements a DC voltage $\sim$2 mV was
applied to the right Ohmic contact and DC current through  
the dot was measured using a DL Instruments model 1211 current
preamplifier connected to the left Ohmic contact.

\section{Theoretical model}
 
The hole manifold with Bloch angular momentum $J=3/2$ is composed of
two HH states, with $J_z=\pm 3/2$, and two LH states, with $J_z=\pm  
1/2$.  
This manifold is described by the Luttinger-Kohn Hamiltonian, which
provides the HH (LH) single-band energy in the anisotropic form  
$\hat{T}_{HH(LH)}=(\gamma_1\pm\gamma_2)(p_x^2+p_y^2)/2m_0
+(\gamma_1\mp2\gamma_2)p_z^2/2m_0+ U(x,y) + V_H(z)$.  
Here $\vec{p}$ is the momentum operator, $m_0$ is the free electron
mass, $U(x,y)$ is the lateral confinement generated by gates, and  
$V_H(z)$ is the vertical confinement produced by the heterointerface
and the top accumulation gate.  
The anisotropic effective masses are defined by the Luttinger
parameters (for GaAs, $\gamma_1=6.95$ and $\gamma_2=2.25$) producing  
the hole subband masses $m_{HH}(z)=0.41m_0$ and $m_{LH}(z)=0.09m_0$.  
As a consequence, the tight vertical confinement $V_H(z)$ results in a
splitting of the hole subbands, with the lowest-lying HH subband lying
lower in energy than the lowest-lying LH subband. 
This splitting is large enough to justify the further development of
our model in the basis of HH states only, with the influence of LH
subbands included perturbatively~[\onlinecite{sbogan}].
Following Bulaev and Loss [\onlinecite{sBulaev}] we write the
perturbative Hamiltonian for a single confined HH as   
\begin{equation} 
\hat{H}=\frac{1}{2m}(p_x^2+p_y^2) + U(x,y) +  
\frac{1}{2}g^* \mu_B B_z \sigma_z + H_{SO}, 
\label{shamil}
\end{equation} 
with the effective spin-orbit (SO) interaction  
$$ 
H_{SO} = i\alpha E_{\perp}(\sigma_+ p_-^3 - \sigma_- p_+^3 ) 
+\beta (\sigma_+ p_- p_+ p_- - \sigma_- p_+ p_- p_+). 
$$ 
Here, the in-plane momentum operator 
$\vec{p}=-i\hbar\nabla +\frac{e}{c}\vec{A}$
contains the magnetic-field dependent term with $e$ being the hole
charge, $c$ the speed of light, and $\vec{A}$ the magnetic vector
potential, such that the field $\vec{B} = \nabla\times \vec{A}$.
In the following we assume  $\vec{B}=[0,0,B_z]$ to be perpendicular to
the heterointerface, which results in the Zeeman term (the third term
in the above Hamiltonian) scaled by the effective hole g-factor
$g^*$, with $\mu_B$ being the Bohr magneton.
We also introduce the effective two-level spin operator $\sigma_z$
such that $J_z = \frac{3}{2}\sigma_z$ [\onlinecite{sBulaev}].
The two other (in-plane) components of that operator define the spin
raising (lowering) operator $\sigma_{\pm}=(\sigma_x\pm i \sigma_y)/2$
appearing in $H_{SO}$ together with $p_{\pm} = p_x \pm i p_y$.
The SO Hamiltonian accounts for Rashba and Dresselhaus interactions  
characterized respectively by parameters $\alpha$ and $\beta$
[\onlinecite{sSzumniak,sDorozhkin,sWinkler}].
The Rashba term is additionally scaled by the effective electric field
$E_{\perp}$ produced by the accumulation gate. 

As evident from the Hamiltonian (\ref{shamil}), the SO interaction
couples the hole spin and orbital degrees of freedom, which leads to
the rotation of the hole spin as it tunnels through the device.    
This tunneling channel appears in addition to the usual,
spin-conserving channel, involving the simple tunneling of the hole
across the interdot potential barrier created electrostatically by the
gates (this barrier is a component of the lateral confinement
$U(x,y)$).
To bring out these two effects clearly, and to enable quantitative
fitting to experimental data, we now map the Hamiltonian (\ref{shamil})
onto a two-site Hubbard model along the procedure described in
Ref.~[\onlinecite{sIrene}].
To this end, we choose the basis in the form of four spin-orbitals
(spinors) $\{ |L,\uparrow\rangle, |L,\downarrow\rangle,
|R,\uparrow\rangle, |R,\downarrow\rangle\}$, where $L$ ($R$) labels
the orbital part being centered on the left (right) dot, while the
arrow denotes the hole spin.
We do not specify the explicit functional form of these spinors, but
choose them to be orthogonal, which eliminates the need of accounting
for the overlap matrix.
We can now generate the matrix elements of our Hamiltonian against
this localized basis.
The diagonal elements, i.e., the energies of each basis state,
attain the form $\langle L,\uparrow| H | L,\uparrow\rangle =
\varepsilon_L + \frac{1}{2}E_Z $, and analogously for the three other
spinors.
The onsite energy $\varepsilon_L$ is a free parameter, tunable by gate
voltages.
The analogous parameter for the right-hand dot orbitals is denoted as
$\varepsilon_R$. 
The spin-conserving tunneling process is enabled by the matrix element
$\langle L,\uparrow| H | R,\uparrow\rangle 
= \langle L,\downarrow| H | R,\downarrow\rangle = -t_N$.
Because of the conservation of spin, this element is computed without
any contribution of $H_{SO}$, and therefore has the same meaning as
that used to describe lateral gated electron systems
[\onlinecite{sIrene}].
The spin-flipping process, on the other hand, is quantified by
$\langle L,\uparrow| H | R,\downarrow\rangle
=-i|t_F|e^{-i\phi}$.
Owing to the form of $H_{SO}$, this element is a complex number, with
the magnitude $|t_F|$ and phase $\phi$, the latter dependent on the
relative strength of the Rashba and Dresselhaus SO terms.
We have found that the theoretical results in this work are
insensitive to the exact value of $\phi$ and in the following
we choose $\phi=0$ for simplicity. 

As a result of the mapping, we end up with the Hubbard Hamiltonian
\begin{equation}
\hat{H} = \left[
\begin{array}{ccccc}
\varepsilon_L + E_Z / 2 & 0 & -t_N & -i t_F  \\
0 & \varepsilon_L - E_Z/2 & -i t_F  & -t_N \\
-t_N & i t_F  & \varepsilon_R +E_Z / 2 & 0 \\
i t_F  & -t_N & 0 & \varepsilon_R - E_Z / 2 \\
\end{array}
\right].
\label{hubhamil}
\end{equation}
identical to the form shown in the main text.
The values of the parameters $t_N$ and $t_F$ and the hole g-factor
$g^*$ are obtained by fitting to the experimental data. 

\section{Theoretical simulations of the experimental system}

\subsection{Coherent and incoherent processes in single-hole tunneling}

As described in the main text, we study the coherent dynamics of our
system in magneto-transport spectroscopy by measuring the tunneling
current as a function of the detuning
$\Delta\varepsilon=\varepsilon_R - \varepsilon_L$ and the magnetic
field.
The sinusoidal microwave modulation for the
Landau-Zener-St\"uckelberg-Majorana interferometry is 
applied to the gate located close to the right-hand dot.
We account for it by replacing 
$\varepsilon_R \rightarrow \varepsilon_R + V_0 \sin(2\pi f t)$, 
where $V_0$ and $f$ are respectively modulation
amplitude and frequency.

\begin{figure}[t]
\includegraphics[width=0.65\textwidth]{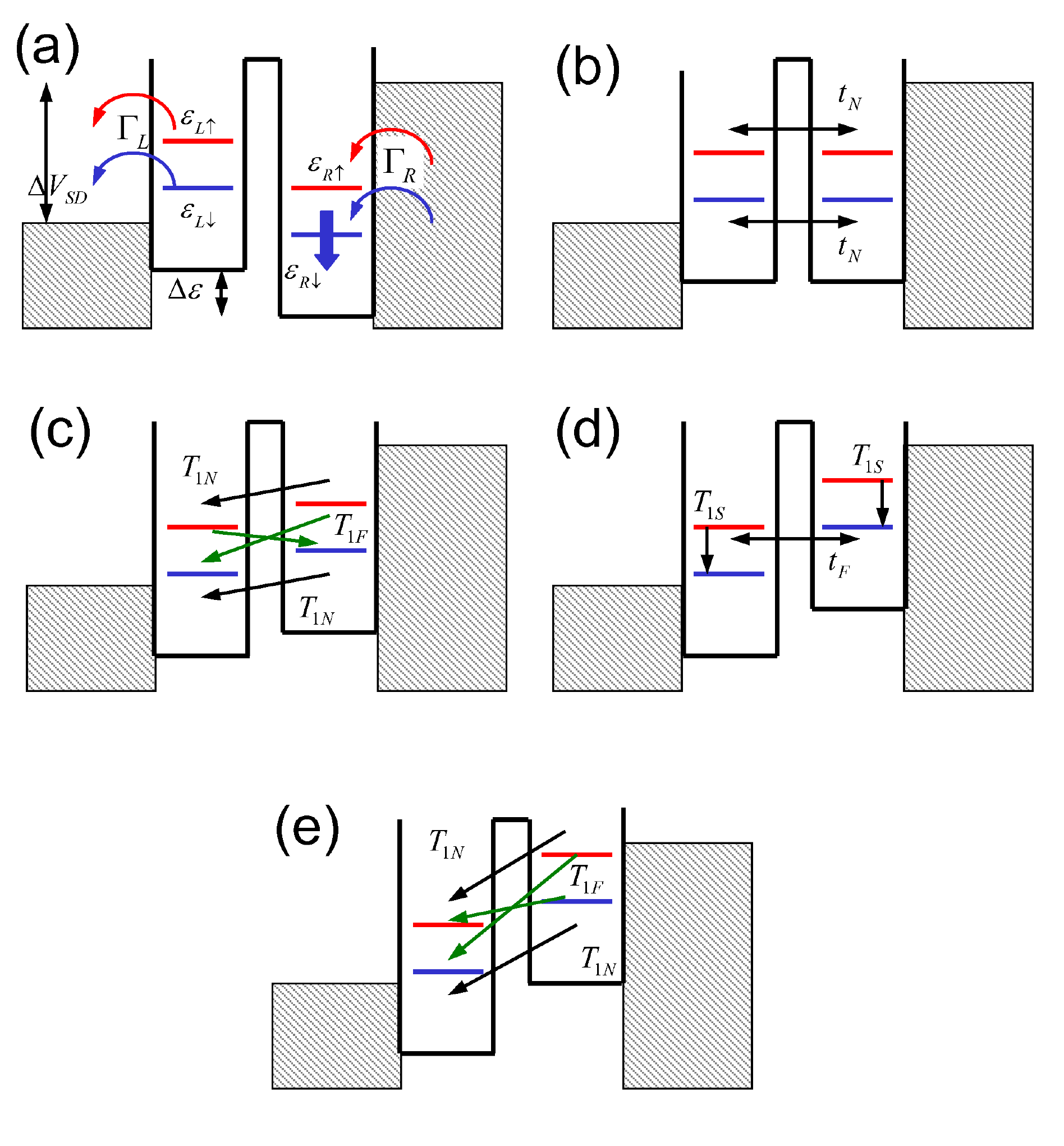}
\caption{Schematic energy diagrams at large source-drain
voltage $V_{SD}$ and with different detunings
$\Delta\varepsilon<0$ (a), $\Delta\varepsilon = 0$ (b),
$0 < \Delta\varepsilon<E_Z$ (c), $\Delta\varepsilon = E_Z$ (d), and
$\Delta\varepsilon > E_Z$ (e).
Double arrows denote resonant tunneling processes, while single arrows
denote leakage channels, spin-conserving (black) and spin-flipping (green).}
\label{sfig1}
\end{figure}

Figure~\ref{sfig1}(a) shows schematically the alignment of levels at 
$\Delta\varepsilon < 0$.
The hole tunnels from the right lead and into the left lead with
spin-independent tunneling rates $\Gamma_R$ and $\Gamma_L$,
respectively. 
The rates are chosen to be high, so that the tunneling current is
determined by the transparency of the interdot barrier.
Due to the large source-drain voltage $\Delta V_{SD} = 2$ mV, both
right-dot spinors are populated equally.
Without microwave modulation, in the alignment shown in
Fig.~\ref{sfig1}(a) the current is blockaded, since the hole is trapped
on the state $|R\downarrow\rangle$ (blue arrow).
Current maxima are expected at detunings $\Delta\varepsilon = 0$
(Fig.~\ref{sfig1}(b)) and $\Delta\varepsilon = E_Z$ (Fig.~\ref{sfig1}(d))
corresponding to the spin-conserving and spin-flipping tunneling,
respectively.
The decoherence of these processes is characterized by times $T_{2N}$
and $T_{2F}$, respectively (not shown in diagrams).
At intermediate ($0 < \Delta\varepsilon < E_Z$, Fig.~\ref{sfig1}(c))
and large detunings ($\Delta\varepsilon > E_Z$, Fig.~\ref{sfig1}(e))
transport can occur only by incoherent phonon-assisted leakage
processes, described by charge relaxation times $T_{1N}$ for the
spin-conserving channel (black arrows), and $T_{1F}$ for the
spin-flipping channel (green arrows). 
Depending on the detuning, these processes can partially cancel out
(panel (c)) or add up (panel (e)) producing different net leakage
current. 
The third type of relaxation process involves the spin flip within one
dot, and is characterized by the time $T_{1S}$ shown schematically
by the vertical arrows in Fig.~\ref{sfig1}(d).

\subsection{Density-matrix approach}

In our theoretical model we account for all coherent and
incoherent processes in the density-matrix rate equation
approach~[\onlinecite{sshevchenko_theory,splatero_theory,splatero2}],
in which we calculate the time-averaged current as a function of 
the detuning, magnetic field, and microwave power.
The quantum master equation for the density matrix $\varrho(t)$
has the form
\begin{equation}
{d\over dt} \varrho(t) = - {i \over \hbar }[H, \varrho(t)] 
+ \Gamma_{in}\varrho(t) + \Gamma_{out}\varrho(t) + \Gamma_{T2}\varrho(t) 
+ \Gamma_{T1}\varrho(t) + \Gamma_{T1S}\varrho(t).
\end{equation}
The density matrix 
\begin{equation}
\rho(t) = \sum_{i,j=1}^{4} \rho_{ij} |i\rangle\langle j|
\end{equation}
is defined in the same four-spinor basis as the Hamiltonian
(\ref{hubhamil}). 
In the master equation the first term on the right-hand side
describes the coherent time evolution of the system (i.e., tunneling).
The other terms describe the tunneling of a single hole from the
right-hand lead into the right-hand dot (the second term), into the
left-hand lead from the left-hand dot (the third term), the
decoherence process (the fourth term), and the $T_1$-type relaxation
processes (the fifth and sixth term).  

In modeling the tunnel coupling to the leads we follow the
Born-Markov secular approximation described, e.g., in
Refs.~\onlinecite{sJouravlev_PRL,sGallego_JAP}.
As already mentioned, owing to the large source-drain bias we assume
that the tunneling rates of both hole spin species are equal and
characterized by the rates $\Gamma_{R}$ for the right-hand lead and
$\Gamma_{L}$ for the left-hand lead.
Modeling of the filling of the right-hand dot involves modifying the
appropriate diagonal density matrix elements subject to the condition
that the total population of the double dot must not exceed one hole.
On the other hand, the emptying of the left-hand dot is modeled by
modifying both diagonal and offdiagonal elements, i.e., we account for
contributions to both incoherent exponential population decay and
decoherence processes. 

The decoherence processes are modeled by suppressing the offdiagonal
density matrix elements connecting single-hole configurations which
in the Hamiltonian (\ref{hubhamil}) are linked by a coherent
tunneling element $t_N$ or $t_F$.
For example, for the spin-flip tunneling process we have
$\langle L,\downarrow|\Gamma_{T2}\varrho(t)|R,\uparrow\rangle =
- \langle L,\downarrow|\varrho(t)|R,\uparrow\rangle / T_{2F}$,
and for the non-spin-flip process
$\langle L, \downarrow | \Gamma_{T2}\varrho(t) |  R,\downarrow \rangle =
- \langle L, \downarrow | \varrho(t) |  R,\downarrow \rangle / T_{2N}$.

The $T_1$-type relaxation processes are modeled in a fashion
similar to the incoherent coupling to the leads, i.e., one state, 
e.g., $|R,\uparrow\rangle$, is depopulated at a constant rate
$1/T_{1F}$ while another, e.g., $|L,\downarrow\rangle$ is populated at
the same rate, obeying the overall charge conservation rule.
The depopulation of the ``source'' level also contributes to all
decoherence processes involving that level.
This example describes the incoherent spin-flip leakage through the
barrier, but, as depicted in Figs.~\ref{sfig1}(c) and (e), we also
account for the spin-conserving leakage characterized by the time
$T_{1N}$. 
Finally, as shown in Fig.~\ref{sfig1}(d), we also account for the
spin-flip relaxation process characterized by the time $T_{1S}$.
The orbitals connected by this process belong to the same
quantum dot. 

We integrate the master equation numerically using the four-point
Runge-Kutta approach.
This allows us to compute the complete density matrix $\varrho(t)$ at
any moment $t$.
The average current integrated  over the total time
of $T_{total} = 5$ $\mu$s is expressed as
\begin{equation}
\bar{I} = {e\Gamma_{L}\over T_{total}} 
\int_0^{T_{total}} dt  [ \varrho(1,1)(t) + \varrho(2,2)(t) ] ,
\end{equation}
i.e., an integral over the time-dependent populations of levels
$|L,\uparrow\rangle$ and $|L,\downarrow\rangle$.

\subsection{Fitting to experimental magnetotransport spectra}

\begin{figure}[t]
\includegraphics[width=0.7\textwidth]{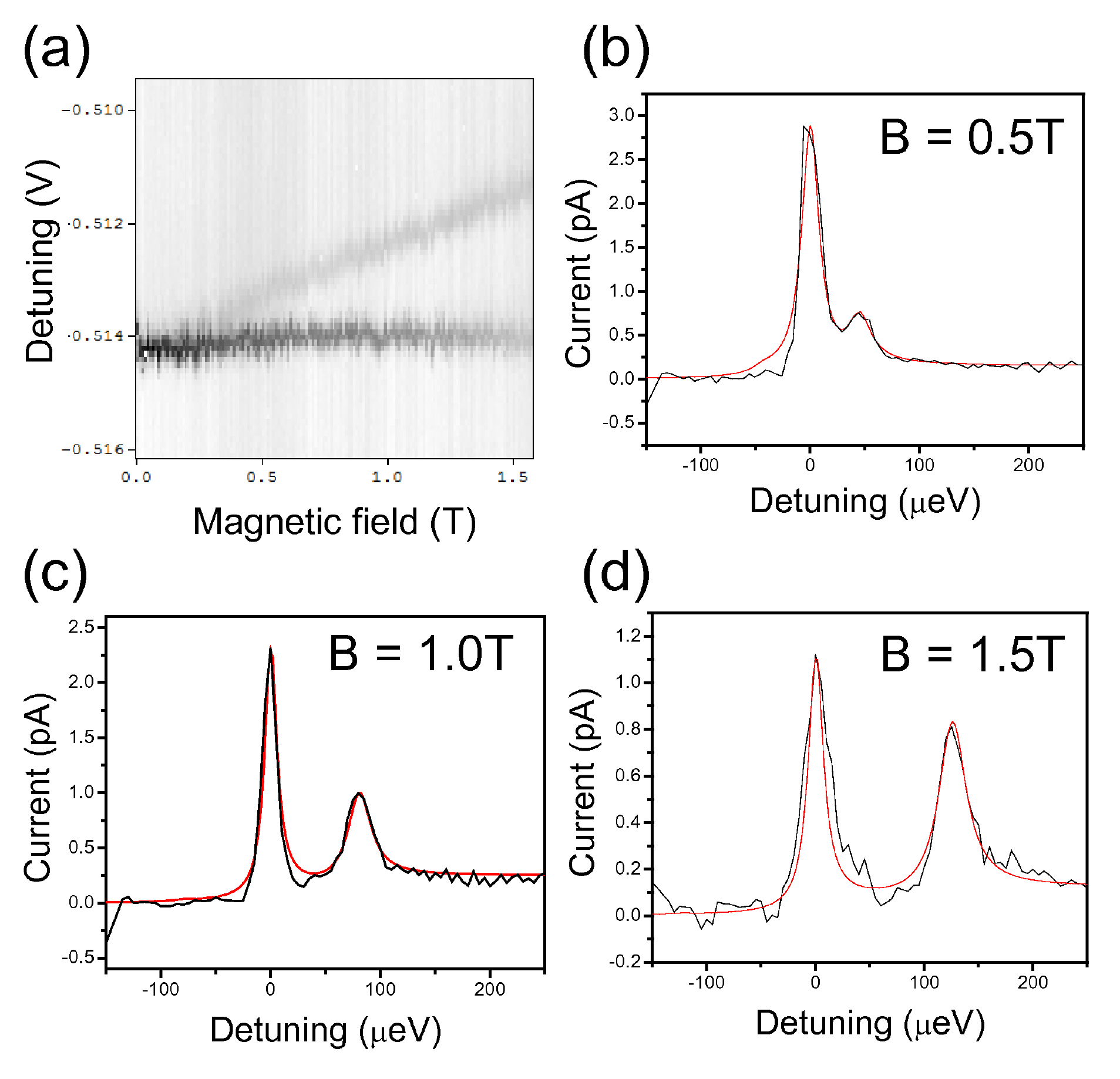}
\caption{(a) Tunneling current measured as a function of the voltage on gate
$R$ (vertical axis) and magnetic field (horizontal axis).
Panels (b), (c), and (d) show the current as a function of detuning at
$B=0.5$ T, $1$ T, and $1.5$ T, measured (black
line) and fitted theoretically (red line).}
\label{sfig2}
\end{figure}

To quantify our theoretical model, we map out the tunneling current as
a function of the static detuning $\Delta\varepsilon$ and the magnetic
field in the magnetotransport experiment. 
The map, shown in Fig.~\ref{sfig2}(a), reveals two resonant features.
The detuning corresponding to one of them is approximately independent
of the field, while the detuning corresponding to the other one
increases linearly with the field.
Based on our model, we identify the first resonance as the 
spin-conserving resonant tunneling (at $\Delta\varepsilon=0$,
Fig.~\ref{sfig1}(b)), and the second one as the
spin-flipping tunneling (at $\Delta\varepsilon=E_Z$,
Fig.~\ref{sfig1}(d)). 
Figures~\ref{sfig2}(b), (c), and (d) show the current extracted
from the panel (a) as a function of the detuning (rescaled by the
lever arm of $50$ $\mu$eV$/$mV) at $B=0.5$ T, $1$ T, and $1.5$ T,
respectively (black lines).  
We extract all model parameters by producing the theoretical fits (red
lines).
By fitting the magnitude and width of the current maxima,
corresponding to the resonant processes, we extract simultaneously
the tunneling elements $t_N$ and $t_F$, the
tunneling rates from/to leads $\Gamma_L=\Gamma_R=2$ GHz,
the decoherence times $T_{2N}\approx T_{2F} = 80$ ps, the spin-flip
relaxation time $T_{1S} = 0.75$ $\mu$s, and the g-factor $g^*=1.35$.
The value of $T_{1S}$ is consistent with the estimates obtained for a
double-dot device in a many-hole regime~\cite{shamilton_gaas}. 
The value of $g^*$ is similar to our previous result for the two-hole
system~\cite{sbogan}.
Finally, the leakage times $T_{1N}=2.7$ $\mu$s and $T_{1F}=2.4$ $\mu$s 
are extracted from the low-current regions at intermediate
(Fig.~\ref{sfig1}(c)) and large detunings (Fig.~\ref{sfig1}(e)).
The values of $t_N$ and $t_F$ obtained in this procedure are
plotted  as a function of the magnetic field in Fig. 1(f) of
the main text (red and black dots, respectively).

\subsection{Dependence of tunneling elements on the magnetic field}

Based on the form of the Hamiltonian (\ref{shamil}) we can
predict the trends of the dependence of both tunneling matrix 
elements $t_N$ and $t_F$ on the magnetic field.
To this end, we define the orbitals centered on the left and right dot
in the model Gaussian form:
$\langle r|L\rangle = \frac{\sqrt{2}}{l\sqrt{\pi}} \exp\left( 
-\frac{(x+d)^2+y^2}{l^2} \right)$ and
$\langle r|R\rangle = \frac{\sqrt{2}}{l\sqrt{\pi}} \exp\left( 
-\frac{(x-d)^2+y^2}{l^2} \right)$.
With this choice the dots lie on the $x$ axis and their centers are
separated by the distance $2d$.
The Gaussian form of the single-particle wave function corresponds to
that of the lowest-energy state of the two-dimensional harmonic
oscillator~\cite{sArek}.
Here $l$ is the characteristic length defining the lateral extent of
the orbitals.
In the presence of the perpendicular magnetic field,
$\frac{1}{l^2} = \frac{m}{\hbar}\sqrt{ \omega_0^2
+\frac{1}{4}\omega_c^2}$, where
$m$ is the in-plane effective hole mass, $\hbar\omega_0$ is the
characteristic frequency of our model harmonic confinement, and 
$\omega_c = \frac{eB}{mc}$ is the cyclotron frequency.
As we can see, with the increase of the magnetic field the
characteristic length $l$ decreases, i.e., our localized orbitals
contract.
The last step in the preparation of our basis is the orthogonalization
of the single-dot orbitals.
We achieve this to first order in the Loewdin procedure by redefining
our basis orbitals as $|\bar{L}\rangle = |L\rangle -
\frac{1}{2}S|R\rangle$
and $|\bar{R}\rangle = |R\rangle - \frac{1}{2}S|L\rangle$
with $S=\exp\left(-2\frac{d^2}{l^2}\right)$ being the overlap of the
original orbitals.

We now proceed to investigating the trends expected for the elements
$t_N$ and $t_F$.
For the former, we assume the interdot barrier to be in the Gaussian
form, $V(x)=V_0\exp\left(-\frac{x^2}{b^2}\right)$ with $V_0$ being the
potential height and $b$ describing the spatial extent of the barrier
along the $x$ axis.
We obtain
\begin{equation}
t_N = \langle \bar{L}| V(x) | \bar{R} \rangle
=V_0 \exp\left(-2\frac{d^2}{l^2}\right) \sqrt{\frac{2b^2}{l^2 + 2b^2}}
\left(1 -  \exp\left(-2\frac{d^2}{2b^2+l^2}\right) \right).
\end{equation}
Following Ref.~[\onlinecite{sbogan}], we take the model values of
parameters $\hbar\omega_0 = 0.29$ meV, $m=0.11 m_0$, $d=57.75$ nm.
The best fit (red dashed line) to the experimental data (red symbols)
plotted in Fig. 1(f) of the main text  is obtained by using the model
barrier with $b=0.5d$ and $V_0=5$ meV.
Since our model treatment does not account for the microscopic detail
of the lateral confinement, such as the anisotropy of the individual dot
potentials and the exact shape of the tnneling barrier,
we do not expect an exact fit.
However, our estimate agrees with the experiment in predicting the
monotonic decrease of $t_N$ as the magnetic field grows, due mainly to
the decrease of the overlap between the orbitals (the first
exponential factor in the above formula).

Let us now move on to the calculation of the element $t_F$.
As is evident from the form of $H_{SO}$, here we need to compute an
integral of generalized momentum in the third power against our model
orbitals.
We choose the magnetic vector potential in the form $\vec{A} =
\frac{B}{2}[ -y, x, 0]$ giving us the uniform magnetic field directed
along the $z$ axis.
We note that, in principle, the localized orbitals $|L\rangle$ and
$|R\rangle$ are defined with a local gauge, whose origin lies at
coordinates $[-d,0,0]$ and $[+d,0,0]$, respectively.
However, the Loewdin procedure mixes these two gauges, and for our
model momentum we choose the global gauge with origin at $[0,0,0]$.
To account for both Dresselhaus and Rashba terms we need to compute
two momentum integrals, i.e., $\langle \bar{L} | p_- p_+ p_- |\bar{R}\rangle$
and $\langle \bar{L} | p_-^3 |\bar{R}\rangle$.
As a result, we obtain the following complex matrix element:
\begin{equation}
\langle \bar{L}\uparrow |H_{SO}|\bar{R}\downarrow\rangle
= - 8\hbar^3 \alpha E_{\perp} \frac{d^3}{l^6} 
\exp\left(-2\frac{d^2}{l^2}\right)
-i\beta\hbar^3 \left( \frac{8d}{l^4}
- \frac{8d^3}{l^6} + \frac{d}{2l_B^4} \right)
\exp\left(-2\frac{d^2}{l^2}\right),
\end{equation}
with the magnetic length $l_B = \sqrt{\frac{\hbar}{m\omega_c}}$.
In analogy to the element $t_N$, here we recover the exponential term
related to the overlap of the two orbitals, indicating that
the magnitude of $t_F$ will decrease exponentially at large fields.
However, unlike in $t_N$, the prefactors generated by the generalized
momentum {\em increase} polynomially as the field grows.
Thus, the resultant trend of the magnitude of $t_F$ is the result
of an interplay between the increasing magnitude of the generalized
momentum and the decreasing overlap between orbitals, producing an
initial increase of $t_F$ at small fields, followed by an exponential
decrease for larger fields.
This is exactly the trend predicted in Fig. 1(f) of the main text
(black dashed line) and seen experimentally (black symbols).
The fit of our model relationship to the experimental
values was achieved by taking $\alpha E_{\perp} = \beta = 133.71$ meV.

\section{Quantitative description of LZSM magneto-tunneling spectra}

\begin{figure}[t]
\includegraphics[width=0.7\textwidth]{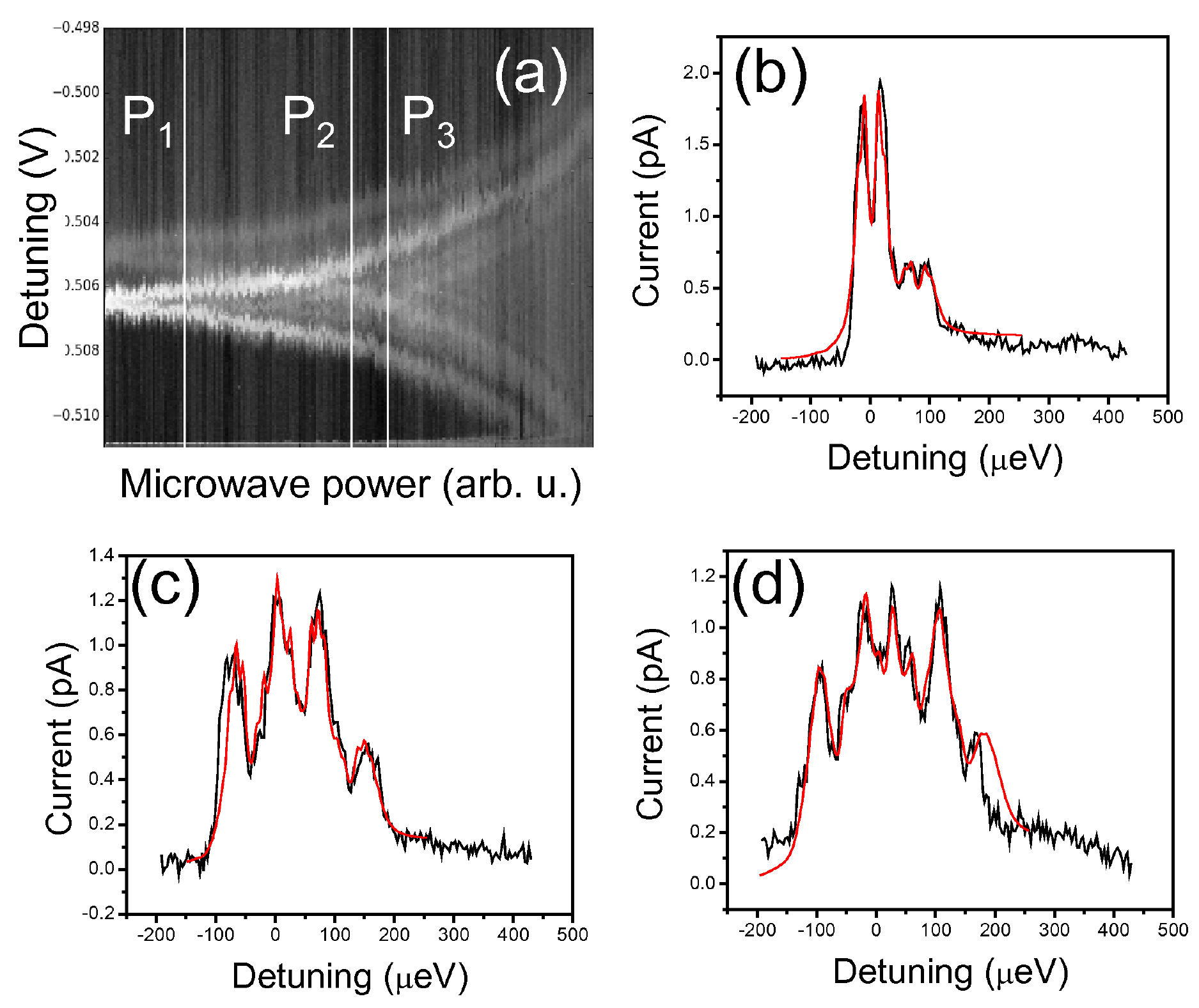}
\caption{(a) Tunneling current measured as a function of the voltage on gate
$R$ (vertical axis) and microwave power (horizontal axis, log scale)
at the magnetic field $B=1$ T.
The microwave frequency $f=2.77$ GHz.
Panels (b), (c), and (d) show the current as a function of detuning at
the three microwave powers, marked in the panel (a) as $P_1$, $P_2$,
and $P_3$, respectively.
Black lines show the measured tunneling current, while red lines show
the theoretical fit.}
\label{sfig3}
\end{figure}

Figure~\ref{sfig3}(a) shows the tunneling current
measured as a function of the microwave power at frequency $f=2.77$
GHz and the magnetic field $B=1$ T.
This panel shows the unprocessed measurement results used to generate
Fig. 3(e) of the main text.
As discussed in the main text, the maxima evolve from the two-peak
structure at low power, through broader, but separate interference
patterns at intermediate powers, towards a complex, overlapping
structure at high powers. 
In Fig.~\ref{sfig3}(b), (c), and (d) we compare the measured trace to
the theoretical fit at powers denoted in panel (a) respectively by
$P_1$, $P_2$, and $P_3$. 
The LZSM driving introduces a precise timing metric, which allows to
fit the tunneling elements $t_N$ and $t_F$ independently from the
decoherence times $T_{2N}$ and $T_{2F}$.
The values of tunneling elements obtained in fitting are
$t_N= 0.255$ $\mu$eV and $t_F=0.205$ $\mu$eV for the power $P_1$,
$t_N= 0.26$ $\mu$eV and $t_F=0.28$ $\mu$eV for the power $P_2$,
$t_N= 0.28$ $\mu$eV and $t_F=0.31$ $\mu$eV for the power $P_3$.
The theoretical fits to the measured current are of high quality for
all microwave modulation powers.

As shown in Figs. 2 and 3 of the main text, at high microwave
frequencies $f$ the interference pattern consists of a set
of interference fringes separated  in detuning by $2\pi\hbar f$.
Only one set is observed at $B=0$ (Fig. 2 of the main text), while
at a nonzero field we see two sets, due to the spin-conserving and
spin-flipping photon-assisted tunneling, respectively.
Our theoretical fits, presented in the right-hand panels of Figs. 2
and 3 of the main text, were generated in the full density-matrix
simulations.
However, the large spacing between maxima allows
to employ a perturbative regime, in which the power dependence of the
peak amplitude of the $n$-th fringe is scaled by the Bessel function 
$J_n(V_0 / 2\pi \hbar f)$ ($n=0,\pm1,\dots$), with $V_0$ being the
microwave amplitude~[\onlinecite{sshevchenko_theory}].
These functions exhibit characteristic oscillations, with 
the central maximum ($n=0$) being the only fringe visible at low
powers. 
This dependence enables tuning of the effective tunneling elements
with microwave power, and activate or deactivate the photon-assisted
tunneling pathways, which are otherwise available based on matching of
detuning, Zeeman energy, and the energy of the microwave photon.

\end{document}